\documentclass[letterpaper,12pt]{extarticle}
\usepackage{graphicx}
\usepackage{dcolumn}
\usepackage{bm}
\usepackage{tabularx} 
\usepackage{preamble}
\usepackage{breqn}
\usepackage{overpic}
\usepackage{circuitikz}
\usepackage{upgreek}
\usepackage{enumerate}
\usepackage{url}
\usepackage{bbm}
\usepackage{setspace}
\usepackage[font=small,labelfont=bf]{caption}
\usetikzlibrary{calc}
\usepackage{extarrows}
\usepackage{amsmath,amssymb}
\usepackage{booktabs,amsfonts,dcolumn}
\usepackage{circledsteps}
\usepackage{subcaption}
\usepackage{booktabs,subcaption,amsfonts,dcolumn}

\numberwithin{equation}{section}
\newcolumntype{d}[1]{D..{#1}}
\definecolor{refkey}{rgb}{0.9451,0.2706,0.4941}
\definecolor{labelkey}{rgb}{0.9451,0.2706,0.4941}
\usepackage{cite}

\def\z2{$\mathbb{Z}_2$}

\newcommand\T[1]{\mathbb{T}^{#1}}
\definecolor{darkgray}{rgb}{0.33, 0.33, 0.33}
\usepackage{braket}

\newcommand{\Tr}{{\mathrm{Tr}}}

\usepackage[left=30mm,right=30mm,top=30mm,bottom=30mm]{geometry}
\usepackage{authblk}
\numberwithin{equation}{section}

\begin{document}
\onehalfspacing
\title{
$6d$ Large Charge and $2d$ Virasoro Blocks
}

\author[1,2]{Jonathan J. Heckman\footnote{\href{mailto:jheckman@sas.upenn.edu}{jheckman@sas.upenn.edu}}}
\author[3]{Adar Sharon\footnote{\href{mailto:asharon@scgp.stonybrook.edu}{asharon@scgp.stonybrook.edu}}}
\author[4]{Masataka Watanabe\footnote{\href{mailto:max.washton@gmail.com}{max.washton@gmail.com}}}
\affil[1]{\it \small Dept. of Physics \& Astronomy, University of Pennsylvania, Philadelphia, PA 19104, USA}
\affil[2]{\it \small Dept. of Mathematics, University of Pennsylvania, Philadelphia, PA 19104, USA}
\affil[3]{\it \small Simons Center for Geometry and Physics, SUNY, Stony Brook, NY 11794, USA}
\affil[4]{\it \small Graduate School of Informatics, Nagoya University, Nagoya 464-8601, Japan}
\date{}
\pagenumbering{gobble}

\maketitle

\thispagestyle{empty}
\begin{abstract}
\noindent
    We compute observables in the interacting rank-one $6d$ $\mathcal{N}=(2,0)$ SCFT at large $R$-charge.
    We focus on correlators involving $\Phi^n$, namely symmetric products of the bottom component of the supermultiplet containing the stress-tensor. By using the moduli space effective action and methods from the large-charge expansion, we compute the OPE coefficients $\braket{\Phi^n\Phi^m\Phi^{n+m}}$ in an expansion in $1/n$. The coefficients of the expansion are only partially determined from the $6d$ perspective, but we manage to fix them order-by-order in $1/n$ numerically by utilizing the $6d/2d$ correspondence. 
    This is made possible by the fact that this $6d$ observable can be extracted in $2d$ from a specific double-scaling limit of the vacuum Virasoro block, which can be efficiently computed numerically. 
    We also extend the computation to higher-rank SCFTs, and discuss various applications of our results to $6d$ as well as $2d$.
\end{abstract}
\newpage
\pagenumbering{arabic}
\pagestyle{plain}
\onehalfspacing
\tableofcontents
\newpage

\newcommand\mc[1]{\mathcal{#1}}
\newcommand\N{{\mathcal{N}}}
\renewcommand\L{{\mathcal{L}}}
\newcommand\todo[1]{\textcolor{red}{(#1)}}
\newcommand\beq{\begin{equation}}
\newcommand\eeq{\end{equation}}

\setcounter{tocdepth}{2}

\section{Introduction}

Superconformal field theories (SCFTs) in six dimensions are non-Lagrangian, and so they are inherently strongly coupled in the sense that there is no obvious small parameter that can be used to perform calculations in perturbation theory. Since they initially appeared using arguments from string theory \cite{Witten:1995zh,Strominger:1995ac,Seiberg:1996qx}, calculations in these theories focused on either observables which are protected by SUSY or calculations in the large-$N$ limit through their holographic dual. 
As a result, it is not clear how to approach computations of various physical quantities at finite $N$ in general.

One of the few available methods in such cases is the large-charge expansion \cite{1505.01537,1611.02912,2008.03308},\footnote{See also \cite{2007.07262,Heckman:2020otd, 2208.02272,Bergman:2020bvi} for related work on large $R$-charge operators of BMN-type \cite{hep-th/0202021} in certain $6d$ $\mathcal{N}=(1,0)$ SCFTs.} which studies generic strongly-coupled conformal field theories with a continuous global symmetry $G$. One can then compute observables (operator dimensions, OPE coefficients, etc.) related to operators with large charge $Q\gg 1$ (or more generally, large representations of $G$).
More concretely, the method uses the state-operator correspondence to map such operators to large-charge states, which are described by an effective field theory (EFT) around a large vacuum expectation value (VEV).
The EFT terms are eventually organized in terms of the $1/Q$-expansion, leading to an emergent weak-coupling parameter $1/Q$ that allows one to extract physical quantities in a simple and controlled fashion.

Further simplifications occur when the theory has a moduli space, which will be the focus of this paper. In particular, at large VEV the effective action is expected to be free,
in contrast to the case without moduli (even in the presence of SUSY), where the leading order EFT is already interacting  \cite{1706.05743,1710.07336,1804.01535,1803.00580,1809.06280,1908.10306,2001.06645,2005.03021,2008.01106,2103.05642,2103.09312,2405.19043,2406.19441,2408.07391}.
There are also far fewer effective operators at each order in the inverse charge expansion.
For example, subleading $F$-terms are completely absent for $4d$ $\mathcal{N}=2$ rank-one SCFTs, as a result of which the OPE coefficients among chiral ring operators are completely fixed by the $a$-anomaly \cite{1804.01535}.

In this paper, we initiate the study of $6d$ SCFTs using the large-charge expansion, expecting that it also provides a new and systematic tool to study various physical quantities in such notoriously non-Lagrangian theories.
We will mostly focus on the interacting $\mathcal{N}=(2,0)$ supersymmetric SCFT with rank one (the $A_1$ theory), and the relevant symmetry will be the $SO(5)$ $R$-symmetry.
We will see that the large-charge analysis is quite useful and in fact produces new results where conventional methods such as holography are not available.

The main tool of our large-charge analysis is the moduli space effective action.
Various symmetries of the problem and the simplicity of the moduli space structure severely constrain the EFT.
In fact, it was argued in \cite{Maxfield:2012aw,Cordova:2015vwa} that the EFT for the rank-one $\mathcal{N}=(2,0)$ theory 
is in principle completely fixed to six-derivative order up to one coefficient, the $a$-anomaly.
However, it is not so easy to determine the explicit form of the bosonic effective action in practice -- as we explain, the arguments of \cite{Cordova:2015vwa} leave one unknown bosonic four-derivative coefficient, whose value is in principle fixed by SUSY but is still unknown. Our analysis will provide the tools to fix this coefficient numerically;
one of our results is therefore an explicit but conjectural expression for the four-derivative bosonic effective action, which turns out to be the four-derivative term of a Dirac-Born-Infeld (DBI) type action.

The EFT we construct describes low-lying operators at large $R$-charge. 
For the $A_1$ theory, in particular, the lowest-dimension operators at fixed $R$-charge live in a half-BPS ring freely generated by the bottom component of the stress-tensor multiplet $\Phi^{IJ}$, which lives in the rank-two traceless symmetric representation of $SO(5)_R$
(here $I,J=1,\dots,5$ are the indices for the vector representation of $SO(5)_R$).
Then, the lowest-dimension operator at large $R$-charge is given by the symmetric product of $\Phi^{IJ}$, which we denote as\footnote{In this normalization we have the chiral ring relation $\Phi^n \times \Phi^m \sim \Phi^{n+m} + \cdots$.}
\begin{align}
    \Phi^n\equiv \Phi^{(I_1J_1}\cdots \Phi^{I_nJ_n)}-(\text{traces})\;.
    \label{eq:normalization}
\end{align}
$\Phi^n$ is in the rank-$2n$ symmetric representation of $SO(5)_R$, and as a half-BPS operator its dimension is fixed by SUSY to be $4n$.
The low-lying spectrum around the large $R$-charge vacuum then describes the low-lying operators above $\Phi^n$.

We will concentrate on two main classes of observables.
First, we compute the dimension of $\Phi^n$ using the large-charge expansion.
This is not very exciting since we already know that it has a protected dimension $4n$, but having no corrections to its dimension is a non-trivial consistency check from the EFT perspective.
The other interesting observable we study is the two-point function of $\Phi^n$,
\begin{align}
\label{eq:6d-twopoint}
    \lambda_n\equiv \abs{x}^{8n}\braket{\Phi^n(x){\Phi}^n(0)}.
\end{align}
As we review around equation \eqref{eq:norm_OPE}, $\lambda_n$ determines the OPE coefficients $\braket{\Phi^n\Phi^m{\Phi}^{n+m}}$ when we work in the standard normalization where $\Phi^n$ has unit two-point function, as opposed to the normalization \eqref{eq:normalization}. While the dimension of $\Phi^n$ is protected by SUSY to be $4n$, the computation of $\lambda_n$ is a much more nontrivial matter.
By using the large-charge EFT, we compute $\lambda_n$ at asymptotically large $n$. We find
\begin{align}\label{eq:lambda_res}
    \log\lambda_n= \log\Gamma(2n+1)+An+(8-160\pi^{3/2}\tilde{b})\sqrt n-2\log n+B+O(n^{-1/2})\;,
\end{align}
where $A$ and $B$ are scheme- and normalization-dependent quantities \cite{1710.07336,1804.01535} and $\tilde{b}\equiv b_1+\frac{b_2}{2}$ is related to the unknown coefficient in front of the effective four-derivative EFT term described above.\footnote{This paragraph, updated in version four on arXiv, corrects an error in which we wrongly concluded that $b_1$, defined in \eqref{eq:4_ders}, vanishes instead of $\tilde{b}$. The error arose in truncating \eqref{eq:O1O2full} to the one-Cartan expression \eqref{eq:S42_reduction}. The affected equations are \eqref{eq:lambda_res}, \eqref{eq:S4_saddle}, \eqref{eq:S4_reduced}, \eqref{eq:OPE}, \eqref{eq:beta_EFT}, \eqref{eq:lambda_conclusion}, \eqref{eq:S42_reduction}, \eqref{eq:S42_decomposition}, \eqref{eq:S42_evaluated}, and \eqref{eq:S4_saddle_appendix}. The error persisted through version three on arXiv and is also in the published version in Physical Review D, for which we publish an erratum.}

The $6d$ perspective alone is thus slightly limited in its power. To proceed further, we utilize the $6d/2d$ correspondence discussed in \cite{Beem:2014kka}. The correspondence maps generators of the ring of half-BPS operators to generators of a $2d$ chiral algebra by restricting them to a $2d$ plane and passing on to the cohomology class defined by certain nilpotent supercharges. In general, $\Phi$ is mapped to the energy-momentum tensor $T$, and for the $A_1$ theory the corresponding $2d$ chiral algebra is just the Virasoro algebra at central charge $c=25$. 

The half-BPS operators $\Phi^n$ are also mapped to elements of this $2d$ chiral algebra. As we review, these elements are given by quasi-primaries $\T{n}$ which are defined as follows: consider some basis for the quasi-primaries $\{ O_i \}$ of given dimension $2n$, written in terms of normal-ordered products of $T$ and its derivatives. Then we can take linear combinations to arrive at a new basis where there is a single quasi-primary $\T{n}$ which includes the term $T^n$, and which is orthogonal to all other quasi-primaries (in the sense that its two-point function with other quasi-primaries vanish). Fixing the coefficient of $T^n$ to be 1 then uniquely determines $\T{n}.$\footnote{As an example, $\T{2}$ is the well-known quasi-primary $T^2-\frac{3}{10}T''$ For other explicit examples see Appendix \ref{app:Tn}.} The $6d/2d$ correspondence maps $\lambda_n$ to the two-point function
\begin{equation}\label{eq:2d_corr}
    \lambda_n=\abs{y}^{4n}\braket{\T{n}(y)\T{n}(0)}\;.
\end{equation}

As we review, this $2d$ observable has many interpretations. For example, it can be understood as a specific element of the inverse of the Kac matrix in the vacuum, or as a coefficient in the large-dimension limit of the vacuum Virasoro block. While $\lambda_n$ can be computed analytically in $2d$ for low $n$, the computation becomes increasingly complicated as $n$ is increased, and the result for general $n$ is not known analytically. Instead, the $2d$ analysis gives us an alternative \textit{numerical} method of computing $\lambda_n$, using the algorithm developed in \cite{Chen:2017yze}.
In particular, we will use the fact that the Zamolodchikov vacuum-exchange $H$-function $H(c,h,q)$ with $h$ the dimension of the identical external operators and $q$ a certain function of the cross-ratio $z$ is given by
\begin{align}
    H(c,h,q)=\sum_{n=0}^{\infty} \frac{(16qh)^{2n}}{\lambda_n}\times \left(1+O(h^{-1})\right)
    \label{eq:H-func}
\end{align}
in the double-scaling limit $q\to 0$ and $h\to\infty$ with $qh$ fixed,
which will be explained and proven in Appendix \ref{sec:vir}.

The $6d$ and $2d$ approaches to computing $\lambda_n$ thus come together beautifully and inform each other. Our $6d$ large-charge analysis allows us to fix the analytic form for the expansion of $\lambda_n$ in orders of $1/n$, except for one undetermined parameter up to the order we are interested in.
The $2d$ numerical computations then allow us to fix this remaining coefficient. The result has new applications for both $6d$ and $2d$:
\begin{itemize}
    \item In the $6d$ $A_1$ theory we are able compute 
    the two-point function $\lambda_n$
    at large $n$ to order $O(n^0)$,
    \begin{align}
        \log \lambda_n=\log\Gamma(2n+1)+An+8\sqrt{n}-2\log n +B+O(n^{-1/2})\;,
    \end{align}
    where, as always, we leave $A$ and $B$ undertermined as being normalization- and scheme-dependent.
    This result allows us to fix the previously unknown four-derivative coefficient in the moduli EFT of the $A_1$ theory, so that the moduli space EFT is completely fixed up to six-derivative terms. 
    \item In $2d$ we are able to compute the two-point function of $\T{n}$ for $c=25$ in an expansion in $1/n$ to high order in $1/n$. 
    As we explained, this can be related to the expansion coefficients of the vacuum Virasoro block in the double-scaling limit mentioned above.
\end{itemize}

This interplay between $6d$ and $2d$ allows us to progress even further.
In terms of $2d$, we can consider the two-point function of $\T{n}$ in CFTs with central charge $c\neq 25$.
This corresponds to higher-rank $6d$ $A_r$ theories as they are mapped to $2d$ theories with central charge $c=4r^3+12r^2+9r$, where $\lambda_n$ again corresponds to the strength of the two-point function of $\Phi^n$.
Whereas the genuine large-charge analysis in $6d$ becomes increasingly complicated for larger rank, we can still use generic large-charge methods to propose a form for the expansion of $\lambda_n$.
In particular, we expect the expansion in terms of $1/n$ to be the same as the $A_1$ theory, but with different coefficients.
We can then fit this result using the $2d$ numerics and fix the coefficients. 
Combining these results from $6d$ and $2d$, we finally conjecture that in the $6d$ $A_r$ theory with $r$ determined by $c=4r^3+12r^2+9r$, we have
\begin{equation}\label{eq:OPE_res}
        \log\lambda_n=\log \Gamma(2n+1)-2n\log 2 +4\sqrt{\frac{c-1}{12}}\sqrt{2n} -\frac{c-1}{12}\log n+B+O(n^{-1/2})\;.
\end{equation}
This can also be interpreted as a general formula regarding the chiral algebra (or more precisely the vacuum Virasoro block) with generic central charge $c>1$, see \eqref{eq:H-func}.
It would be very interesting to reproduce this result directly in $2d$, and we leave this to future work.

Finally, we numerically find that the subleading piece of $\lambda_n$ shows an interesting behavior at or near $c=1$.
We show that it contains a term of the form
\begin{align}
    \log \lambda_n = \cdots + \frac{0.25}{n}\sin\left(8.0\sqrt{n}\right) + \cdots \;,
\end{align}
entirely from $2d$ numerics.
This is important as it can be interpreted as a correction coming from a BPS worldsheet instanton on the moduli space of some unknown $6d$ non-unitary interacting SCFT whose chiral ring corresponds to the $c=1$ chiral algebra.

The outline of this paper is as follows. In section \ref{sec:6d} we set up the calculation by reviewing the $6d$ $\mathcal{N}=(2,0)$ SCFT and the known results for the moduli space EFT on the rank-one tensor branch. We then perform the computation of the OPE coefficients $\lambda_n$ using methods from the large-charge expansion. In section \ref{sec:2d} we use the $6d/2d$ correspondence  to map $\lambda_n$ to a $2d$ observable in a Virasoro chiral algebra with central charge $c=25$. This computation can be done numerically to very high order, and we use it to fix the unknown coefficients at order $O(\sqrt{n})$ in the expansion of $\lambda_n$. We also use these results to fix the previously unknown four-derivative term in the moduli space EFT. In section \ref{sec:higher_rank} we discuss the extension of these results to central charge $c>1$, or equivalently to higher-rank $6d$ $A_r$ theories. We also discuss possibilities of interpreting certain oscillations in data as coming from the BPS string worldsheet on the moduli in terms of $6d$. Finally we conclude and discuss some open questions in section \ref{sec:future}.

\section{\texorpdfstring{$6d$}{6d} \texorpdfstring{$\mathcal{N}=(2,0)$}{N=(2,0)} SCFTs at Large \texorpdfstring{$R$}{R}-Charge}\label{sec:6d}

\subsection{Review of the \texorpdfstring{$6d$}{6d} \texorpdfstring{$\mathcal{N} = (2,0)$}{N=(2,0)} Theories}

We first review some features of the $6d$ $\mathcal{N} = (2,0)$ SCFTs.
The $6d$ $\mathcal{N}=(2,0)$ superconformal algebra in Lorentzian signature is given by $\mathfrak{osp}(6,2|4)$ \cite{Nahm:1977tg}.
Its bosonic subalgebra is $\mathfrak{so}(6,2)\oplus \mathfrak{sp}(4)$, where $\mathfrak{so}(6,2)$ is the conformal algebra and $\mathfrak{sp}(4)\cong\mathfrak{so}(5)$, the $R$-symmetry. At the level of free fields, there is a unique
massless low-spin representation of the superconformal algebra given by the Abelian tensor multiplet. In terms of
component fields, the bosonic content consists of five real scalars transforming in the $\mathbf{5}$ of the $\mathfrak{so}(5)_{R}$ $R$-symmetry and a chiral two-form potential with self-dual field strength.\footnote{One can also consider a CPT conjugate presentation with an anti-chiral two-form potential and anti-self-dual field strength. This is especially common in string constructions.} The fermionic content consists of a pseudo-Majorana-Weyl spinor (i.e., satisfying a symplectic condition) in the $\mathbf{4}$ (the spinor representation) of $\mathfrak{so}(5)_{R}$. Famously, there are no relevant or marginal perturbations of a free $6d$ theory, and so in order to find interacting SCFTs one has to resort to other methods. See references \cite{Heckman:2018jxk, Argyres:2022mnu} for recent reviews of $6d$ SCFTs.

The best evidence for the existence of $6d$ SCFTs comes from string-based constructions. This involves taking a suitable decoupling limit in a local model involving singularities and/or branes. In particular, the $\mathcal{N} = (2,0)$ SCFTs were first constructed in \cite{Witten:1995zh, Strominger:1995ac} and were recognized as ``ordinary'' quantum field theories in \cite{Seiberg:1996qx}. A systematic approach to realizing the $\mathcal{N} = (2,0)$ theories is via type IIB string theory on the background $\mathbb{R}^{5,1} \times \mathbb{C}^2 / \Gamma_{\mathrm{ADE}}$ where $\Gamma_{\mathrm{ADE}}$ is a finite subgroup of $SU(2)$ chosen so as to preserve sixteen real supercharges in the $6d$ spacetime. There is an ADE classification of such finite subgroups, and this is in one to one correspondence with the ADE series of simply laced Lie algebras $\mathfrak{g}_{\mathrm{ADE}}$. This then produces the celebrated ADE classification of such theories.\footnote{Strictly speaking this leads to a collection of relative QFTs since one ends up with a vector of partition functions rather than a single partition function. Such subtleties will not concern us in this work.} One can also realize the $A_r$ series with an additional ``center of mass'' tensor multiplet via $r$ coincident M5-branes. Similar considerations hold for $r$ M5-branes (and their images) in the presence of an O5-brane (see \textit{e.g.}, \cite{Hanany:2000fq}).

For $\mathfrak{g}$ a Lie algebra of ADE type, there is a corresponding $6d$ $\mathcal{N} = (2,0)$ SCFT. We denote by $r_{\mathfrak{g}}$ the rank of the Lie algebra. The tensor branch moduli space is then given by:
\begin{equation}
\mathcal{M}_{\mathfrak{g}} \simeq \mathbb{R}^{5 r_{\mathfrak{g}}} / \mathcal{W}_{\mathfrak{g}}\;,
\end{equation}
where $\mathcal{W}_{\mathfrak{g}}$ is the Weyl group of the Lie algebra. In the specific case of $\mathfrak{g} = A_{r}$ the Weyl group is just $S_{r+1}$, the symmetric group on $r+1$ letters. Observe that in the presentation of the $A_r$ series (including a free center of mass tensor multiplet) in terms of $r+1$ coincident M5-branes, the five directions transverse to the M5-branes geometrically implement the $\mathbb{R}^{5}$ directions. Similar considerations apply in the near horizon limit of a large number of coincident M5-branes, as realized by the holographic dual $AdS_7 \times S^4$ where the $R$-symmetry acts as the isometries of the $S^4$ factor.

Determining the operator content of $6d$ SCFTs remains an outstanding open problem.
There are strong constraints from representation theory of the superconformal algebra
(see \cite{Cordova:2016xhm} and references therein for a general discussion). Our interest here will be in the half-BPS operators. They sit in rank-$k$ traceless symmetric representations of $SO(5)_R$ and have conformal dimensions $\Delta=2k$. In $6d$ SCFTs with a holograhic dual, these can be accessed via the representation theory of the corresponding supergravity backgrounds via references \cite{Gunaydin:1984fk, Gunaydin:1984wc, Gunaydin:1985tc}. A notable example is the bottom component of the stress-tensor multiplet, $\Phi$, which is in the rank-two traceless symmetric representation of $SO(5)_R$.
They are believed to form a half-BPS ring, which is freely generated by elements in one-to-one correspondence with the Casimir invariants of $\mathfrak{g}$ \cite{hep-th/9707079,hep-th/0702069}.
For the $A_1$ theory in particular, the half-BPS ring is freely generated only by $\Phi$.
In other words, its members are given by symmetric products of $\Phi$, denoted by $\Phi^n$ -- they are in the rank-$2n$ traceless symmetric representation of the $R$-symmetry group, and have operator dimension $4n$.
As $\Phi^n$ is the lowest dimension operator at large $R$-symmetry representations, we shall primarily focus on this operator in much of the present work.

\subsubsection{Weyl anomalies}

To better understand the structure of correlation functions involving $\Phi$, it will be helpful to have a more precise characterization of the Weyl anomalies of the $6d$ $\mathcal{N} = (2,0)$ SCFTs.

Recall that the Weyl anomaly is the one-point function of the trace of the stress-tensor in a generic curved background \cite{Capper:1974ic}. Its scheme-independent part can be written as
\begin{align}
    \braket{T_{\mu}^{\mu}}= aE_6+\sum_{i=1}^3 c_iI_i +I_F\;,
\end{align}
where $E_6$ is the six-dimensional Euler density, $I_i$ are six-derivative Weyl invariants, and $I_F$ represents terms due to the background $\mathfrak{so}(5)$ field strength (which we ignore in what follows).
Our normalization for $E_6$ is determined by writing it on conformally flat space using curvature tensors as \cite{1205.3994}
\begin{align}
    E_6
    \equiv \frac{3}{2} R^\mu{ }_\nu R^\nu{ }_\sigma R^\sigma{ }_\mu-\frac{27}{20} R^\mu{ }_\nu R^\nu{ }_\mu R+\frac{21}{100} R^3\;,
\end{align}
whereas the normalization for $I_i$ does not matter in this paper and we just refer to \cite{hep-th/0001041} for definitions.
Importantly, though, $\mathcal{N}=(2,0)$ SUSY is believed to relate $c_i$'s by a single constant \cite{hep-th/0001041,Beem:2014kka}, so we normalize $c\equiv c_1=c_2=c_3$ hereafter, by properly normalizing $I_i$. We shall find it convenient to use a convention which is slightly different from that provided by Bastianelli, Frolov and Tseytlin (denoted as [BFT]) \cite{hep-th/0001041,Beem:2014kka} which we reference as $\left(E_6\right)^{\rm [BFT]}$. In our conventions $\left(E_6\right)^{\rm [BFT]}=-8E_6$.

The $a$- and $c$-anomalies for various $6d$ $\mathcal{N}=(2,0)$ theories have been computed in the literature \cite{Henningson:1998gx, Graham:1999jg, Intriligator:2000eq, hep-th/0001041, Cordova:2015vwa}. We work in a normalization where $a_{U(1)} = c_{U(1)}$ for a free tensor multiplet:
\begin{align}
    a_{U(1)}=c_{U(1)} = -8a_{U(1)}^{\rm [BFT]}=\frac{7}{(4\pi)^3 144}\;.
\end{align}
In general, we have for $\mathfrak{g}$ an ADE type Lie algebra with rank $r_{\mathfrak{g}}$, dimension $d_{\mathfrak{g}}$ and dual Coxeter number $h_{\mathfrak{g}^{\vee}}$:
\begin{align}
    &a_{\mathfrak{g}}=\left(\frac{16}{7}h_{\mathfrak{g}}^{\vee} d_{\mathfrak{g}} + r_{\mathfrak{g}} \right)a_{U(1)}\;, \\
    &c_{\mathfrak{g}}=\left(4 h_{\mathfrak{g}}^{\vee} + r_{\mathfrak{g}} \right)c_{U(1)}\;.
\end{align}
In particular, in the special case of $\mathfrak{g} = A_{r}$, we have:
\begin{align}
    &a_{r}=\left(\frac{16r^3}{7}+\frac{48r^2}{7}+\frac{39r}{7}\right)a_{U(1)}\;, \\
    &c_{r}=(4r^3+12r^2+9r)c_{U(1)}\;.
\end{align}

\subsection{The Moduli Space EFT}

We are interested in the lowest-dimension operators at large representations of $SO(5)_R$ in the $A_1$ theory.
As we have discussed, these are simply given by the large symmetric products of $\Phi$, denoted $\Phi^n$.
We use the idea of the large-charge expansion to study them.

The heart of the large-charge expansion lies in writing down the effective field theory around a large dimensionful VEV which is helical in time \cite{1505.01537}.
In cases with a moduli space such an EFT is nothing but the moduli space effective action, organized by the derivative expansion, analogous to \cite{1706.05743,1804.01535}.
As we have reviewed, the moduli space for the $A_1$ theory is $5$-dimensional and is parameterized by the VEV of the scalar fields $\varphi^{I}$ in an Abelian tensor multiplet, where $I$ is the index of the fundamental representation of $SO(5)_R$.

The moduli space EFT of the tensor branch of the $A_1$ theory has been studied in \cite{Cordova:2015vwa}, as we now review.
We will only concern ourselves with the bosonic effective action hereafter as the rest will not be relevant in the following.
Let us write the bosonic EFT (in Lorentzian signature with mostly plus metric) in terms of the derivative expansion as
\begin{align}
    L_{\rm EFT}=L_{\partial^2}+L_{\partial^4}+L_{\partial^6}+\cdots,
\end{align}
where $L_{\partial^i}$ refers to the effective Lagrangian at $i$-derivative order.
At each order, the effective operators need to respect the superconformal symmetry of the original SCFT and in particular they need to be $R$-symmetry invariant as well as be Weyl-covariant with weight six.

Let us discuss the general structure of this effective expansion according to \cite{Cordova:2015vwa}.
In the following we write down the expressions only on flat space, but we assume that a suitable Weyl-completion is always possible.
In practice this is enough because we will only work on $\mathbb{R}\times S^5$ or $S^6$, which we can reach from flat space \it via \rm Weyl transformations.

\begin{description}

\item[{Two-derivative effective action}]
The leading order effective action is given by the free kinetic term, normalized so that
\begin{align}
    L_{\partial^2}=L_{\rm free}\equiv -\frac{1}{2}\sum_{I=1}^{5}\left(\partial_\mu\varphi^I\right)^2.
\end{align}
The leading order equation of motion (EOM) can be used to simplify organizations of higher-order terms, which on flat space is
\begin{align}
    \partial_\mu\partial^\mu \varphi^I=0\;.
\end{align}

\item[{Four-derivative effective action}]

As argued in \cite{Maxfield:2012aw,Cordova:2015vwa}, there is only one effective operator at four-derivative level.
It was also argued in \cite{Cordova:2015vwa} that the coefficient of the effective operator is completely determined by the theory's $a$-anomaly.
However, the explicit form of these terms is not known.
Luckily, part of the argument in \cite{Cordova:2015vwa} leading to uniqueness allows us to write down two allowed effective operators, $\mathcal{O}_1$ and $\mathcal{O}_2$, at four-derivative level, and uniqueness tells us that only a specific (unknown) combination of the two operators is allowed to appear.
We will fix the relative coefficient using another argument later in Section \ref{sec:numerical}.

We defer the discussions of determining the form of $\mathcal{O}_{1,2}$ to Appendix \ref{sec:o12} and just present the results here.
They are given by
\begin{align}
\label{eq:O1O2full}
    \begin{aligned}
        \mathcal{O}_{1}&\equiv \frac{\partial_\mu \varphi^I \partial^\mu \varphi^I \partial_\nu \varphi^J \partial^\nu \varphi^J}{\abs{\varphi}^3}\sim \frac{\left(\partial^2 \abs{\varphi}^2\right)^2}{4\abs{\varphi}^3}\;,\\
        \mathcal{O}_{2}&\equiv \frac{\partial_\mu \varphi^I \partial^\mu \varphi^J \partial_\nu \varphi^I \partial^\nu \varphi^J}{\abs{\varphi}^3}\sim \frac{\partial^2 (\varphi^I\varphi^J) \partial^2 (\varphi^I\varphi^J)}{4\abs{\varphi}^3}\;,
    \end{aligned}
\end{align}
where $\sim$ refers to an equality modulo the leading order EOM and total derivatives.
The latter expressions are more convenient for conducting Weyl transformations and so we use these expressions hereafter.
To summarize, we have fixed the form of the four-derivative effective action to be
\begin{align}\label{eq:4_ders}
    L_{\partial^4}=b_1\mathcal{O}_1+b_2\mathcal{O}_2=b_1\frac{\left(\partial^2 \abs{\varphi}^2\right)^2}{4\abs{\varphi}^3}+b_2\frac{\partial^2 (\varphi^I\varphi^J) \partial^2 (\varphi^I\varphi^J)}{4\abs{\varphi}^3}\;.
\end{align}
Where $b_1,b_2$ are undetermined coefficients.
For later convenience we define $b\equiv b_1+b_2$ and $\tilde{b}=b_1+\frac{b_2}{2}$, so that
\begin{align}
    L_{\partial^4}=\tilde{b}\mathcal{O}_1+(b-\tilde{b})(2\mathcal{O}_2-\mathcal{O}_1)\;.
\end{align}

Further use of SUSY can determine $b_1+b_2$ in terms of the theory's $a$-anomaly, as it controls the coefficient of the dilaton-only part of the four-derivative interaction in the moduli space EFT \cite{Cordova:2015vwa}.
Let us write $L_{\partial^4}$ in terms of $\psi\equiv \abs{\varphi}$ and other fields representing the Nambu-Goldstone (NG) bosons.
Taking out the dilaton-only part, it becomes
\begin{align}
    L_{\partial^4}\biggr|_{(\partial\psi)^4} =
    (b_1+b_2)\frac{(\partial\psi)^4}{\psi^3}\;.
\end{align}
Now, \cite{Cordova:2015vwa} proved that the coefficient of $\frac{(\partial\psi)^4}{\psi^3}$ obeys
\begin{align}
    b\equiv b_1+b_2=\sqrt{\frac{7}{98304\pi^3}\frac{\Delta a}{a_{U(1)}}}\;,
\end{align}
where $\Delta a\equiv a_1-a_{U(1)}$.
This fixes the overall coefficient for the four-derivative effective action, leaving one relative coefficient between $\mathcal{O}_{1,2}$ to be determined later.
In fact we will see in Section \ref{sec:numerical} that $\tilde{b}\equiv b_1+\frac{b_2}{2}=0$, so that $L_{\partial^4}=b(2\mathcal{O}_2-\mathcal{O}_1)$; together with $L_{\partial^2}$, this is the expansion of the DBI action up to four derivatives.


\item[{Six-derivative effective action}]

Following \cite{1912.09479,2012.10450,2103.13395}, let us schematically decompose the six-derivative effective action as
\begin{align}
    L_{\partial^6}=L_{\rm Euler}+L_{\text{B-type}}+L_{F_{\mu\nu}}+L_{SO(5)_R}+L_{\rm inv}\;,
\end{align}
where $L_{\rm Euler}$ is the Euler action, $L_{\text{B-type}}$ and $L_{F_{\mu\nu}}$ are the terms which produce the $B$-type Weyl anomaly $I_{1,2,3}$ and $I_F$ respectively, $L_{SO(5)_R}$ is the term which accounts for the $SO(5)_R$-anomaly, and $L_{\rm inv}$ represents invariant terms allowed by symmetries.
For the purposes of this paper, it is enough to know the part of this Lagrangian containing the dilaton without derivatives as will become clear from the VEV we are interested in.
Defining the dilaton $\uptau$ by
\begin{align}
    \abs{\varphi}\equiv e^{-2\uptau} \quad \text{or}\quad  \uptau \equiv -\frac{1}{2} \log \abs{\varphi}\;,
\end{align}
such a contribution only comes from $L_{\rm Euler}$ in a form in which the dilaton simply multiplies the Euler density, where the coefficient is the $a$-anomaly difference between the original SCFT and a single Abelian tensor multiplet.
All in all, we have
\begin{align}
    L_{\partial^6}=\Delta a\times \uptau E_6 +\text{(derivatives)}\;,
\end{align}
where $\Delta a\equiv a_{1}-a_{U(1)}$.

\end{description}

To summarize, the part of the effective action we are interested in is
\begin{align}
    \begin{split}
        L_{\rm EFT} = \underbrace{-\frac{1}{2}\sum_{I=1}^{5}\left(\partial_\mu\varphi^I\right)^2}_{\equiv L_2}
        +\underbrace{b_1\frac{\left(\partial^2 \abs{\varphi}^2\right)^2}{4\abs{\varphi}^3}+b_2\frac{\partial^2 (\varphi^I\varphi^J) \partial^2 (\varphi^I\varphi^J)}{4\abs{\varphi}^3}}_{\equiv L_4}\quad\quad\quad\\
    +\underbrace{\Delta a\times \uptau E_6}_{\equiv L_6}+\cdots
    \end{split}
    \label{eq:fulleft}
\end{align}
where $\uptau \equiv -\frac{1}{2}\log \abs{\varphi}$ and $b\equiv b_1+b_2=\sqrt{\frac{7}{98304\pi^3}\frac{\Delta a}{a_{U(1)}}}$.

\subsection{Spectrum of the \texorpdfstring{$A_1$}{A1} Theory at Large Charge}

As we have discussed, the lowest-dimension operator at large $R$-charge is given by $\Phi^n$ which is half-BPS.
Its operator dimension is fixed to be $4n$ by supersymmetry, and so computing its operator dimension from the EFT gives us a nice consistency check of our formalism.

Symmetry identifies the operator $\Phi^n$ with $\varphi^{2n}$ in the EFT (up to normalization), which is a symmetric product of the field $\varphi$ which appears in the EFT.
The idea here, as usual in the large charge expansion, is to use the state-operator correspondence --
we study the energy of the lowest-energy state on $S^5\times \mathbb{R}$ at the representation we are looking at, which is then replaced by a corresponding VEV.
Even though there is a complication due to the presence of multiple commuting charges to turn on, it does not bother us as we are only interested in the traceless symmetric representations.
We can argue from group theory that the corresponding VEV has only one Cartan rotating $\varphi^{1,2}$ turned on, while the other one rotating $\varphi^{3,4,5}$ is turned off \cite{1610.04495,1705.05825,1804.06495,1904.09815,1909.02571,2203.08843,1909.02571,2003.13121}.
All in all, the VEV we are interested in is
\begin{align}
    \left(\varphi^1,\,\varphi^2,\,\varphi^3,\,\varphi^4,\,\varphi^5\right)
    =\sqrt{2}A
    \left(\cos (2t),\,\sin (2t),\,0,\,0,\,0\right),
    \label{eq:vev}
\end{align}
where
\begin{align}
    A^2=\frac{n}{2\mathop{\mathtt{Vol}}(S^5)}\;, \quad \mathop{\mathtt{Vol}}(S^{d-1})=\frac{2\pi^{d/2}}{\Gamma(d/2)}\;.
\end{align}
The helical frequency is fixed by the leading order free EOM of the moduli space EFT, while $A^2$ was determined from the fact that we are in the rank-$2n$ symmetric traceless representation.
For convenience we also define
\begin{align}
    \phi\equiv\frac{1}{\sqrt{2}}\left(\varphi^1+i\varphi^2\right),\quad \phibar\equiv\frac{1}{\sqrt{2}}\left(\varphi^1-i\varphi^2\right)\;,
\end{align}
so that the VEV is given by
\begin{align}
    \phi=Ae^{2it}, \quad \varphi^{3,4,5}=0\;.
\end{align}
Incidentally, we now see that the large charge EFT we wrote down in \eqref{eq:fulleft} is organized in terms of a $1/n$-expansion, up to order $O(\log n)$.

As a consistency check, let us compute the operator dimension of $\Phi^n$ using the EFT up to six-derivative order, which is given by the energy of the configuration given in \eqref{eq:vev} on the cylinder.
Its classical piece is given by simply substituting \eqref{eq:vev} into the Hamiltonian obtained from our EFT, and the quantum corrections are smaller than $O(n^0)$ which we will not analyse in this paper.

\begin{description}
    \item[Two-derivative] The two-derivative part of our EFT on the cylinder is given by
\begin{align}
    L_2\equiv -\frac{1}{2}\sum_{I=1}^{5}\left(\left(\partial_\mu\varphi^I\right)^2+4\left(\varphi^I\right)^2\right),
\end{align}
where $4\left(\varphi^I\right)^2$ is the conformal coupling.
The classical energy evaluated on \eqref{eq:vev} is immediately given by $4n$.
\item[Four-derivative] Let us proceed to $L_4$.
Its flat-space (Euclidean) expression can be conformally transformed to the (Lorentzian) cylinder by using
\begin{align}
    ds^2_{\rm flat}=e^{2it}\left(-dt^2+d\Omega^2_{S^5}\right)
    \equiv r^2ds_{\rm cyl}^2\;,
\end{align}
where the radial coordinate $r$ in $\mathbb{R}^6$ is related to the time direction in $S^5\times \mathbb{R}^5$ as $r\equiv e^{it}$.
The conformal transformation maps a field $F_{\rm flat}$ of weight $k$ on flat space to a counterpart on the cylinder $F_{\rm cyl}$ as
\begin{align}
    F_{\rm flat} = r^{-k}F_{\rm cyl}\;.
\end{align}
The Laplacian $\triangle_{\rm flat}$ acting on $F_{\rm flat}$ can expressed using $F_{\rm cyl}$ as
\begin{align}
        \begin{split}
        \triangle_{\rm flat}F_{\rm flat}
        &=
        r^{-5}\partial_r \left(r^5 \partial_rF_{\rm flat}\right)+r^{-2}\triangle_{S^5}F_{\rm flat}\\
        &=e^{-i(k+2)t}\left(-e^{-i(k-4)t}\partial_t\left\{e^{4it}\partial_t\left[e^{-ikt}F_{\rm cyl}\right]\right\}+\triangle_{S^5}F_{\rm cyl}\right)\\
        &\equiv r^{-k-2}\triangle_{\rm cyl}F_{\rm cyl}\;.
    \end{split}
\end{align}
Now that the expressions for $\mathcal{O}_{1,2}$ are given on the cylinder, it is a simple exercise to plug the VEV in and evaluate the classical pieces.
We indeed see that they vanish, giving no correction to the leading order formula at order $O(\sqrt{n})$.
\item[Six-derivative] Finally, $L_6$ is already vanishing on a cylinder as the Euler density $E_6$ is vanishing.
\end{description}

We therefore conclude from the EFT that the operator dimension of $\Phi^n$ is $4n$ without any corrections up to $O(n^0)$.
This is consistent with the fact that $\Phi^n$ is half-BPS.

The EFT also allows us to study operators of charge $Q$ whose dimension is above the BPS bound $\Delta_{BPS}=4n$. Most interestingly, it allows for a computation of the dimension of non-protected operators as well, see \textit{e.g.},~\cite{1706.05743} for an example. However, the computation is more involved in the $6d$ $\mathcal{N}=(2,0)$ case, since the second lowest operators are all still protected. In fact, operators whose dimension is not protected must have $\Delta> \Delta_{BPS}+6$ \cite{Cordova:2016emh}, so that the computation involves operators much higher above $\Delta_{BPS}$ than in \cite{1706.05743}, which complicates the computation. We hope to return to these computations in the future.

\subsection{OPE Coefficients}\label{sec:OPE}

Let us compute the coefficient of the two-point function of the half-BPS operator $\Phi^n$ from the EFT, defined as
\begin{align}
    \lambda_n\equiv |x|^{8n}\braket{\Phi^n(x) \Phi^n(0)}\;,
\end{align}
in the normalization where the chiral ring relations are
\begin{equation}\label{eq:norm}
    \Phi^n\times \Phi^m = \Phi^{n+m}+\cdots
\end{equation}
The physical meaning of $\lambda_n$ becomes clear if we instead use the standard normalization for the operators, such that their two-point functions have unit coefficient:
\begin{align}
    \Phi^n_{\rm CFT}\equiv \frac{\Phi^n}{\sqrt{\lambda_n}}\;,
    \quad \braket{\Phi^n_{\rm CFT}(0)\Phi^n_{\rm CFT}(x)}=\frac{1}{\abs{x}^{8n}}\;.
\end{align}
Then the OPE coefficient becomes
\begin{equation}\label{eq:norm_OPE}
    \braket{\Phi^n_{\rm CFT}(x_1)\Phi^m_{\rm CFT}(x_2)\Phi^{n+m}_{\rm CFT}(x_3)}=\sqrt{\frac{\lambda_{n+m}}{\lambda_n\lambda_m}}\frac{1}{\abs{x_{13}}^{8n}\abs{x_{23}}^{8m}}\;.   
\end{equation}
From now on we use the normalization \eqref{eq:norm}.

As discussed in the previous subsection, the half-BPS operator $\Phi^n$ corresponds to $\varphi^{2n}$ in EFT terms (up to a normalization), and we only need to evaluate the two-point function of $\varphi^{2n}$.
For example, we can insert $\varphi^{2n}$ on the north and the south pole of the unit $S^6$ and compute $\lambda_n$ using the EFT.
One caveat here is that $\lambda_n$ has some ambiguities in its definition.
This results from the fact that we do not know the relative normalization between $\Phi^n$ and $\varphi^{2n}$ and that the overall normalization for $\lambda_n$ is scheme-dependent because it is given by the sphere partition function.
This results in ambiguities in the $O(n)$ and $O(n^0)$ part of $\log\lambda_n$, and the former will eventually get cancelled when computing the OPE coefficient in the usual CFT sense where all two-point functions are unit normalized.
The latter is in principle calculable after careful matching of renormalization schemes, but we will not pursue in this paper.

Let us now compute $\lambda_n$, following the $4d$ calculation in \cite{1710.07336,1804.01535,Arias-Tamargo:2019xld,Watanabe:2019pdh,Badel:2019oxl}.
The strategy here is to insert $\varphi^{2n}(x_1)$ and $\varphi^{2n}(x_2)$ in the path-integral,
\begin{align}
    \braket{\varphi^{2n}(x_1)\varphi^{2n}(x_2)}
    &=\frac{1}{Z}\int \mathcal{D}\varphi\, 
    \varphi^{2n}(x_1)\varphi^{2n}(x_2)e^{-\int d^6x L_{\rm E}[\varphi]}\;,
    \label{eq:pathint}
\end{align}
where $Z$ is the partition function that normalizes the two-point function.
The geometry on which we place the theory is strictly speaking $S^6$, but most of the computations can be unambiguously done on $\mathbb{R}^6$ by using a conformal transformation. Here we have written down the action in Euclidean signature obtained by Wick rotating \eqref{eq:fulleft} by substituting $t=-i t_{\mathrm{E}}$ and introducing an overall minus sign to the action, which results in
\begin{align}
    \begin{split}
        L_{\rm E} = {\frac{1}{2}\sum_{I=1}^{5}\left(\partial_\mu\varphi^I\right)^2}
        -L_4-L_6-\cdots,
    \end{split}
    \label{eq:fullefte}
\end{align}
where
\begin{align}
    \begin{split}
        L_4&\equiv b_1\mathcal{O}_1+b_2\mathcal{O}_2\equiv b_1\frac{\left(\partial^2 \abs{\varphi}^2\right)^2}{4\abs{\varphi}^3}+b_2\frac{\partial^2 (\varphi^I\varphi^J) \partial^2 (\varphi^I\varphi^J)}{4\abs{\varphi}^3}\;,\\
        L_6 &\equiv \Delta a\times \uptau E_6\;,
    \end{split}
\end{align}
where $\uptau \equiv -\frac{1}{2}\log \abs{\varphi}$.
For later convenience, we denote
\begin{align}
    S_2&\equiv \frac{1}{2}\int d^6x {\sum_{I=1}^{5}\left(\partial_\mu\varphi^I\right)^2}\;,\\
    S_{2i+2}&\equiv \int d^6x\, L_{2i+2} \quad \text{for} \quad i\geq 1\;.
\end{align}
We can then evaluate the path-integral \eqref{eq:pathint} by using the saddle-point approximation at large-$n$, by bringing the insertions up inside the exponential.

It is helpful to stop here for a moment to understand the general structure of the $1/n$-expansion.
First of all, our large-charge effective action itself is ordered in terms of $n^{-1/2}$, such that $S_{2i+2}$ scales as $O(n^{1-i/2})$ modulo an overall $\log n$.
We will then take the leading order saddle-point, just by using $L_2$, and sum over all the vacuum diagrams, taking into account the possible tadpoles if any.
By closely following the arguments in \cite{1710.07336,1804.01535}, we will see that the vacuum diagrams scale as
\begin{align}
    n^{1-L-\sum_{i=1}^{\infty}2^{i-2}N_{2i+2}}\;,
\end{align}
where $L$ is the number of loops and $N_{2i+2}$ is the number of vertices generated from $S_{2i+2}$.
We will depict the $k$-leg vertex generated from $S_{2i}$ as
\begin{align}
        \begin{tabular}{c}
            \includegraphics[width=0.2\linewidth]{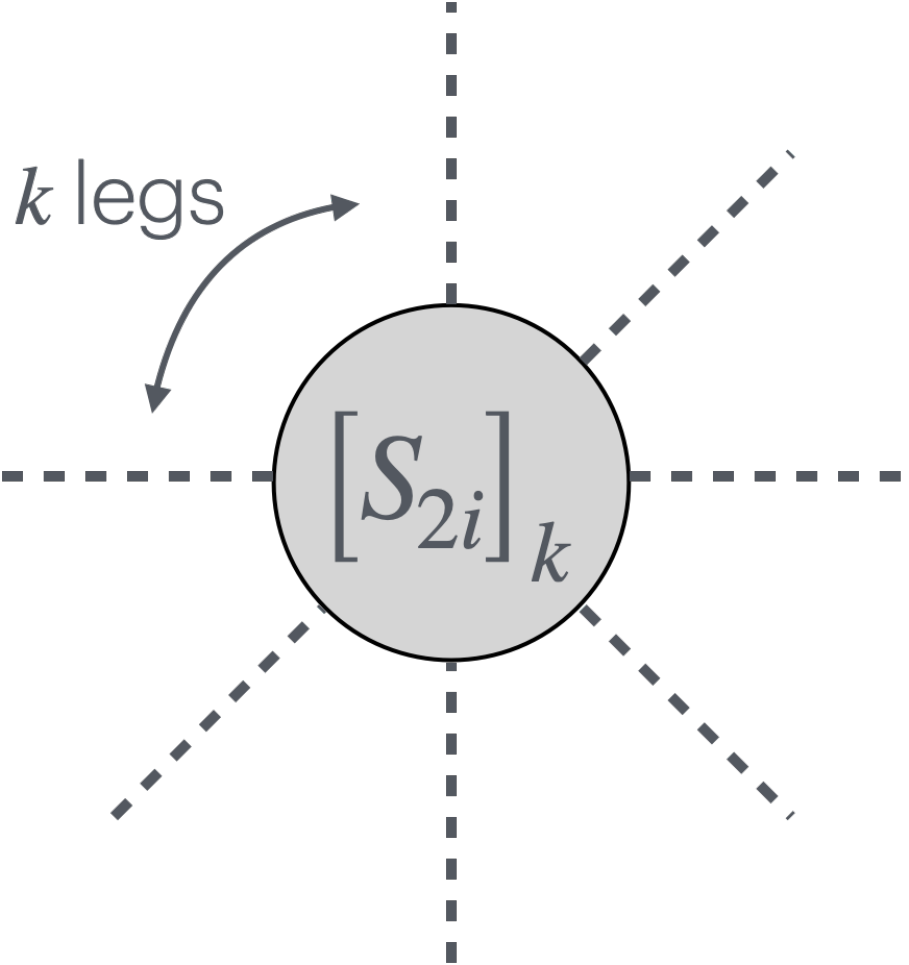}
        \end{tabular}
\end{align}
hereafter.
Note that there can be no-leg vertices (\it i.e., \rm $k=0$), which correspond to $S_{2i}$ evaluated on the saddle-point, which are in general non-vanishing.

We are only interested in computing $\lambda_n$ up to order $O(\log n)$ in this paper, in which case the computation simplifies quite a lot.
The only possible diagrams contributing at $O(n^0)$ or above are \Circled{1} the one-loop diagram without vertices, \Circled{2} the tree diagram with one vertex from $S_4$, \Circled{3} the tree diagram with two vertices from $S_4$, and \Circled{4} the tree diagram with one vertex from $S_6$:
We depict these diagrams in Table \ref{tab:diagrams}, along with their scalings.

\begin{table}[t!]
    \centering
    \begin{tabular}{cccc}
        & diagram & description & scaling \\ \midrule\midrule
        \Circled{1} &  $\vcenter{\hbox{\includegraphics[width=0.1\linewidth]{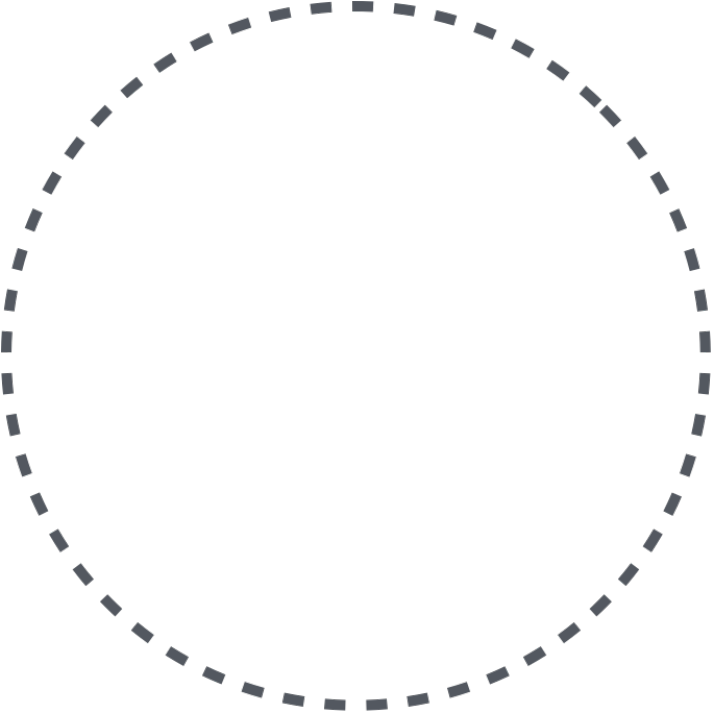}}}$ & one-loop vacuum diagram & $O(\log n)$ \\\midrule
        \Circled{2} & $\vcenter{\hbox{\includegraphics[width=0.05\linewidth]{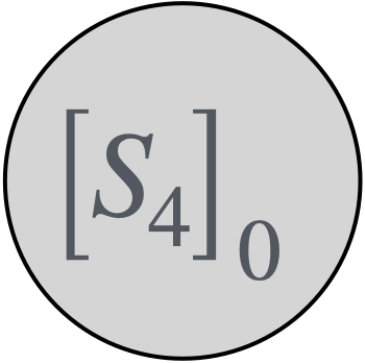}}}$ & no-leg vertex from $S_4$ & $O(n^{1/2})$ \\\midrule
        \Circled{3} & $\vcenter{\hbox{\includegraphics[width=0.22\linewidth]{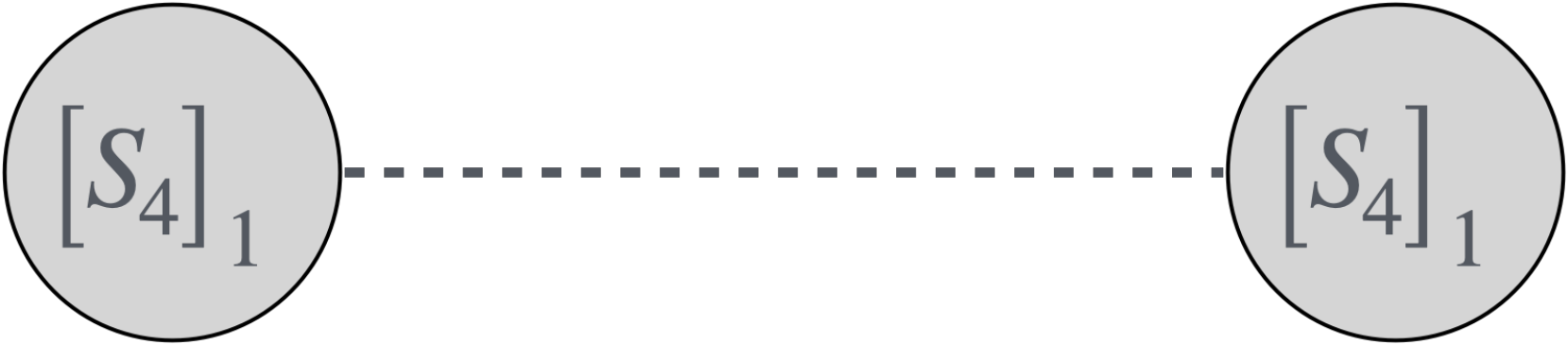}}}$ & tree diagram with one-leg vertices from $S_4$ & $O(n^0)$ \\\midrule
        \Circled{4} & $\vcenter{\hbox{\includegraphics[width=0.05\linewidth]{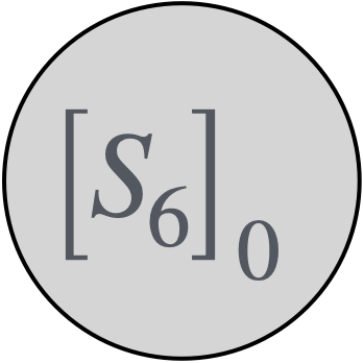}}}$ & no-leg vertex from $S_6$ & $O(\log n)$\\
        \bottomrule
    \end{tabular}
    \caption{Diagrams (potentially) contributing to $\log \lambda_n$ at $O(n^0)$ or higher.}
    \label{tab:diagrams}
\end{table}

Expressed in words, \Circled{1} and \Circled{4} contribute at $O(\log n)$. Additionally, \Circled{2} contributes at $O(n^{1/2})$. These are the only contributions above $O(n^0)$.\footnote{As a tree diagram containing two one-point vertices, \Circled{3} would contribute at $O(n^0)$ if it existed at all -- It in fact does not even exist because $S_4$ does not contain a piece which is linear in fluctuations around our saddle-point.}
In other words, we will only have to evaluate the classical action of $S_4$ and $S_6$ on the saddle-point to account for \Circled{2} and \Circled{4}, while we can utilize Wick's theorem in order to account for \Circled{1}.

\begin{description}

\item[Two-derivative]

The two-derivative effective Lagrangian $L_2$ is simply given by the free kinetic term of $\varphi$.
The contribution to $\lambda_n$ from the effective action at this order is then just given by the Wick contraction, thus we have
\begin{align}
    \log \lambda_n=\log \Gamma(2n+1)+(\text{higher-derivative})\;.
\end{align}
Note that $\log \Gamma(2n+1)$ is an all-orders formula in terms of the saddle-point approximation of $\log \lambda_n$, if the Lagrangian consisted only of $L_2$.
For example, the $O(\log n)$ piece in $\log \Gamma(2n+1)$ comes from the one-loop correction to the classical saddle-point action.

Even though we avoided using the saddle-point configuration by using Wick's theorem, we need it in order to evaluate the four- and six-derivative action at the classical saddle-point.
It is given by minimizing the leading order action $S_2$ with source terms
\begin{align}
    S_{\rm full}\equiv \int d^6 x \left(\abs{\partial\phi}^2-2n \delta^6(x-x_1)\log \phi(x)-2n \delta^6(x-x_2)\log \phibar(x)\right)\;,
\end{align}
where we have already used the fact that we can set $\varphi^{3,4,5}=0$ at the saddle-point, as in \eqref{eq:vev}.
We have also replaced $\varphi^{1,2}$ with $\phi$ and $\phibar$, where the overall normalization of the insertions was not taken care of as it contributes to the $O(n)$-part in $\log \lambda_n$ and simply is convention-dependent as discussed.
The result for the leading-order saddle-point is then given by solving the EOM,
\begin{align}
    \partial^\mu\partial_\mu\phi=-\frac{2n}{\phibar}\delta(x-x_2)\;,\\
    \partial^\mu\partial_\mu\phibar=-\frac{2n}{\phi}\delta(x-x_1)\;,
\end{align}
whose solution becomes
\begin{equation}
\label{eq:6d_sol}
    \phi(x)=\frac{e^{i \theta_0}x_{12}^2}{\sqrt{4\pi^3}(x-x_2)^4} (2n)^{1 / 2}\;, \quad \phibar(x)=\frac{e^{-i \theta_0}x_{12}^2}{\sqrt{4\pi^3}(x-x_1)^4} (2n)^{1 / 2}\;,   
\end{equation}
where $\theta_0$ is some undetermined parameter, representing the degrees of freedom of rotating $\phi$ in the complex plane.
Our saddle-point spontaneously breaks (not only $U(1)\subset SO(5)_R$ as expressed by $\theta_0$ but also) the $SO(5)_R$ symmetry and so we would have to recover it by orbit averaging as proposed in \cite{1612.08985,1706.05743,2103.16580}, but we do not discuss this as it contributes to $\lambda_n$ only at $O(n^0)$ or below.

\item[Four-derivative]

As we have discussed, we only need to plug the leading-order saddle-point given in \eqref{eq:6d_sol} into $S_4$.
We compute this in Appendix \ref{app:four-derivative} and only show the result here, 
\begin{align}\label{eq:S4_saddle}
    S_4\biggr|_{\rm saddle}=(96b_1+176b_2)\pi^{3/2}\sqrt {n}\;.
\end{align}
By using the relation
\begin{align}
    b\equiv b_1+b_2 = \sqrt{\frac{7}{98304\pi^3}\frac{\Delta a}{a_{U(1)}}}=\frac{1}{32\pi^{3/2}}\;,\quad \tilde{b}\equiv b_1+\frac{b_2}{2}\;,
\end{align}
we can also write
\begin{align}\label{eq:S4_reduced}
    S_4\biggr|_{\rm saddle}=\left(8-160\pi^{3/2}\tilde{b}\right)\sqrt{n}\;.
\end{align}

\item[Six-derivative]

We only need to plug the leading-order saddle-point given in \eqref{eq:6d_sol} into $S_6$.
We compute this in Appendix \ref{app:four-derivative} and only show the result here, 
\begin{align}
    S_6\biggr|_{\rm saddle}=-2\log n\;.
\end{align}

\end{description}

Summing up all the contributions to $\lambda_n$
we have our final result (note the \it positive \rm sign of $S_{4,6}\bigr|_{\rm saddle}$, they are \it minus \rm the Euclidean action and hence contribute \it positively \rm to $\log \lambda_n$)
\begin{align}\label{eq:OPE}
    \begin{split}
        \log \lambda_n&=\log \Gamma(2n+1)  + S_4\biggr|_{\rm saddle}+S_6\biggr|_{\rm saddle}+ An+B+O(n^{-1/2})\\
        &=\log \Gamma(2n+1) + An+\left(8-160\pi^{3/2}\tilde{b}\right)n^{1/2}-2\log n+B+O(n^{-1/2})\;.
    \end{split}
\end{align}
We will fix $\tilde{b}$ later on to be vanishing from numerics, and will also fix $A$ and $B$.

\subsubsection{Absence of higher-derivative corrections}

We now discuss how higher-derivative corrections affect our result \eqref{eq:OPE}. In the case of a single (2,0) tensor multiplet, the higher-derivative terms were classified in \cite{Cordova:2016xhm}, and our discussion follows \cite{Cordova:2015vwa}. There are two types of terms:
\begin{enumerate}
    \item $F$-terms $\mathcal L_F=Q^8(\Phi^{(I_1}...\Phi^{I_n)}-\text{traces})$ with $Q$ the supercharge. Importantly, $(\Phi^{(I_1}...\Phi^{I_n)}-\text{traces})$ is half-BPS and so $\mathcal L_F$ involves only 8 supercharges, and as a result only 4 derivatives. 
    \item $D$-terms $\mathcal{L}_D=Q^{16} \mathcal{O}$ which contain at least 8 derivatives.
\end{enumerate}
We thus immediately learn that higher-derivative terms are necessarily $D$-terms. We can thus use the standard argument to show that these $D$-terms do not contribute to correlators of half-BPS operators.\footnote{Schematically, the argument is as follows. Splitting the supercharges into $Q_i,\bar Q_i$ with $i=1,\dots,8$, consider a half-BPS operator $\bar Q_i\mathcal{O}=0$. In perturbation theory, $D$-term contributions to correlators of $\mathcal{O}$ take the form $\braket{\mathcal{O}\cdots \mathcal{O}(\int Q^8\bar Q^8 \Phi)^n }$. Using the fact that for a conserved charge
$\sum_i \braket{ \phi_1\cdots(Q\phi_i)\cdots\phi_n }=0$
and the fact that $\bar Q_i^2=0$ and $\bar Q_i \mathcal{O}=0$, we find that these contributions vanish.} In particular, the two-point function we computed above does not receive corrections from these higher-derivative terms. 

We thus learn that in principle, the OPE coefficients can be computed using only the EFT we wrote above by including all quantum corrections. Indeed, for 4d rank-one theories a similar argument was used to compute the corresponding OPE coefficients exactly to all orders in $1/Q$ \cite{1804.01535}. Unfortunately the $4d$ computation requires additional input which is not accessible in $6d$; specifically, the $4d$ correlators are related by recursion relations, which are derived from differential equations with derivatives taken with respect to an exactly marginal operator. While the same analysis is impossible in $6d$ (due to the absence of exactly marginal operators), hopefully some other input will be enough to fix the result. We leave this to future work.

\section{\texorpdfstring{$6d/2d$}{6d/2d} Correspondence}\label{sec:2d}

$6d$ $\mathcal{N} = (2,0)$ SCFTs are known to have a sector of operators isomorphic to a $2d$ chiral algebra \cite{Beem:2014kka}.
There is a one-to-one correspondence between the generators of the chiral algebra and the generators of the ring of half-BPS operators, by restricting them to a fixed $2d$ plane (which we set to the $x_1$-$x_2$ plane without loss of generality) and passing on to the cohomology class defined by certain nilpotent supercharges.
In particular, it was argued in \cite{Beem:2014kka} that the chiral algebra corresponding to a $6d$ $\mathcal{N}=(2,0)$ SCFT labeled by the Lie algebra $\mathfrak{g}$ is the $\mathcal{W}$-algebra of type $\mathfrak{g}$. 

In this section we map our $6d$ discussion of OPE coefficients above to a calculation of OPE coefficients in a $2d$ chiral algebra (see \cite{Beem:2014kka} for a similar calculation, albeit for a different set of operators and at large $N$). This $6d$/$2d$ mapping then teaches us about both $6d$ and $2d$ theories, since it will allow us to combine the analytic approach from $6d$ with a numerical approach in $2d$ to fix the expansion of the coefficients $\lambda_n$ to high order in $1/n$.

\subsection{The \texorpdfstring{$6d/2d$}{6d/2d} Correspondence for the \texorpdfstring{$A_1$}{A1} Theory}
\label{app:6d2d}

Our main focus in this paper has been the $A_1$ SCFT, whose associated $2d$ chiral algebra is simply the Virasoro algebra with $c=25$.
We are in particular interested in the bottom component of the stress-tensor multiplet $\Phi$ and the symmetric products $\Phi^n$ thereof. Since the $2d$ chiral algebra is just the Virasoro algebra, $\Phi^n$ must map to some combination of products of $T$ and their derivatives, which we denote by $\T{n}$. This combination must be a quasi-primary of dimension $2n$. The explicit expression for the $\T{n}$ has already been derived from analogous discussions using the $4d$/$2d$ correspondence \cite{Beem:2013sza}, see \cite{Beemtalk}, which we now briefly review.

First, we can immediately identify $\T{1}=T$ as the only quasi-primary with dimension 2. Next, let us determine $\T{2}$. Again there is a unique quasi-primary at this order, but let us instead appeal to another argument which is more easily generalized to higher orders. $\T{2}$ has dimension $4$, so it is a linear combination of $T^2$ and $\partial^2 T\equiv T^{\prime\prime}$
(unless otherwise stated all operators are normal ordered).
The latter trivially corresponds to the $6d$ operator $\partial_{12}^2\Phi$, where $\partial_{12}$ generates rotations in the $x_1-x_2$ plane. Since $\Phi^2$ and $\partial^2\Phi$ are orthogonal (in the sense that their two-point functions vanish), $\T{2}$ must be orthogonal to $T''$. Up to an overall normalization we therefore find $\T{2}=T^2-\frac{3}{10}T^{\prime\prime}$. As a consistency check, this is indeed the unique quasi-primary of dimension 4.  

We can generalize this procedure to find $\T{n}$ for general $n$. An important additional point that is required in order to isolate $\T{n}$ is the ``triangle inequality'' \cite{Beem:2017ooy,Beemtalk}. Schematically, this states that an operator composed of $m$ $\Phi$'s is mapped to some combination of $T$'s and derivatives where each term has at most $m$ $T$'s. We can then define $\T{n}$ as follows. First we choose a basis of all quasi-primaries at this order. We can go to a new basis where there is a single basis element which includes the term $T^n$, and normalize the coefficient of $T^n$ to be 1. Finally, we use the Gram-Schmidt algorithm to generate a new orthonormal basis of quasi-primaries, where the term $T^n$ appears in a single basis element with coefficient 1. This basis element is precisely $\T{n}$. We present the expression for $\T{n}$ for small values of $n$ in Appendix \ref{app:Tn}.

Having defined $\T{n}$, the mapping tells us that the two-point functions of $\T{n}$ and $\Phi^n$ match:
\begin{equation}
    \lambda_n=|x|^{8n}\braket{ \Phi^n(x)\Phi^n(0) }=|y|^{4n}\braket{ \T{n}(y)\T{n}(0) } \;,
\end{equation}
where we emphasize that $x$ is a $6d$ coordinate while $y$ is a $2d$ coordinate. We thus turn our attention to the $2d$ two-point function 
\begin{equation}
    \braket{\T{n}(y)\T{n}(0)}=\frac{\lambda_n}{|y|^{4n}}\;.
\end{equation}
Note that we have the OPE $\T{n}\times\T{m}=\T{n+m}+...$, and so if we normalize $\T{n}$ to have unit two-point function then we can use $\lambda_n$ to read off the three-point function of $\T{n}$'s.

The coefficient $\lambda_n$ has many interpretations in $2d$. For example, one interpretation (see Appendix \ref{app:Tn}) is as a specific coefficient of the inverse of the Kac matrix:
\begin{equation}
    \lambda_n=\frac{1}{[M^{-1}]^{\{2^n\},\{2^n\}}}\;,
\end{equation}
where $M_{\{k\},\{k'\}}=\braket{ L_{\{k\}}L_{-\{k'\}} }$ is the Kac matrix evaluated in the vacuum, $L_{\{k\}}=L_{k_1}...L_{k_m}$ are products of the generators of the Virasoro algebra and we use the notation
\begin{equation}
    \{2^n\}=\{\overbrace{2,2...,2}^n\}\;.
\end{equation} 
In the next section we discuss an alternative interpretation which is more suitable for computations.

\subsection{Numerical Computation of the OPE Coefficient}

\subsubsection{Setup}

The $2d$ computation of $\lambda_n$ is hard at large-$n$ in practice; the orthogonalization procedure quickly goes out of hand as we increase the level because the number of candidate operators grows quickly.
Luckily $\lambda_n$ can be read off from a large-dimension limit of the Virasoro conformal block, for which a fast numerical algorithm using the Zamolodchikov recursion relation is known \cite{Chen:2017yze}.

Deferring the explanation to Appendix \ref{sec:vir}, we simply state the relation between the Virasoro conformal block and $\lambda_n$.
Let us concentrate on the Zamolodchikov $H$-function $H(c,h,q)$ which is defined in \eqref{eq:vir}.
This is related to the vacuum Virasoro block with internal dimension $0$ and where we have identical external operators of dimension $h$.
The relation between $\lambda_n$ and $H(c,h,q)$ is as follows: By taking the double-scaling limit $q\to 0$ and $h\to \infty$ with $qh$ fixed, we have
\begin{align}
    H(c,h,q)=\sum_{n=0}^{\infty}\frac{(16qh)^{2n}}{\lambda_n}\times (1+O(h^{-1}))\;.
\end{align}
This is very useful because \cite{Chen:2017yze} provides a fast numerical algorithm to compute the expansion coefficients $d_{2n}$ of $H(c,h,q)$ in terms of $q$,
\begin{align}
    H(c,h,q)\equiv \sum_{n=0}^{\infty} d_{2n}q^{2n}\;,\quad 
    d_{2n}=(\text{polynomial of $h$ of order $2n$})\;.
\end{align}
We can therefore relate $\lambda_n$ to $d_{2n}$ as
\begin{align}
    \lambda_{n}=\lim_{h\to\infty}\frac{2^{8n}h^{2n}}{d_{2n}}\;.
\end{align}

Note that our double-scaling limit corresponds to a certain thermodynamic limit of CFT, as discussed in \cite{1710.10458}.
It is largely unexplored compared to that of large central charge, which has been studied in the context of ${\rm AdS}_3/{\rm CFT}_2$ \cite{1986JETP...63.1061Z,cmp/1103941860,1501.05315,1502.07742,1510.00014,1609.07153,1804.06171,1806.04352,1910.04169}.
It is also not to be confused with the regime of large \it internal \rm dimensions, where, along with the regime of large central charge, an analytic use of the Zamolodchikov recursion relation is possible \cite{1986JETP...63.1061Z,cmp/1103941860,2007.10998,Cardona:2020cfy}.
See also \cite{1711.09913,1810.01335,1905.02191} for generic large-order behavior of the Virasoro block in $2d$ CFTs.

Because of the double-scaling limit we take, we modified the program given in \cite{Chen:2017yze} which implemented the Zamolodchikov recursion relation to compute $\lambda_n$.
The original algorithm computes the $2n$-th coefficient $d_{2n}$ by using the Zamolodchikov recursion relation \cite{Zamolodchikov:1987avt,cmp/1103941860}.
Our modification is so that we only take the leading order in $h$ in each step of the recursion relation --
We modify $R_{m,n}$ given in (A.7) in \cite{Chen:2017yze} to
\begin{align}
    R_{m,n}\approx\frac{8\lambda^2\prod_{p,q}\left(\lambda_{p,q}\right)^2}{\prod_{k,\ell}^{\prime}\left(\lambda_{k,\ell}\right)^2}\;,
    \label{eq:modifiedRmn}
\end{align}
where $c\equiv 13+6\left(b+\frac{1}{b}\right)$, $h\equiv \frac{1}{4}\left(b+\frac{1}{b}\right)-\lambda^2$, and $\lambda_{m,n}\equiv \frac{1}{2}\left(\frac{m}{b}+nb\right)$
(see \cite{1502.07742} for more complete explanation, such as index ranges of the products). 
Because of the limitation of the (original as well as the modified) algorithm, we are not able to set $c=25$ exactly.
This is because $R_{m,n}$ will have a pole at $b=1$ (\it i.e., \rm $c=25$), even though they should cancel in the final result when summing up $R_{m,n}$ as in (A.9) of \cite{1502.07742}.
We instead use $c=24.999$ in order to compute $\lambda_n$, and we checked that the results are stable upon slightly changing $c$.

\subsubsection{Numerical results}\label{sec:numerical}

Let us recap what we expect of $\lambda_n$ from the EFT analysis in $6d$ (see \eqref{eq:OPE}):
\begin{align}
    \log \lambda_n&=\log \Gamma(2n+1) + An+\beta n^{1/2}+\alpha\log n+B+O(n^{-1/2})\;,
\end{align}
where
\begin{align}\label{eq:beta_EFT}
    \beta=8-160\pi^{3/2}\tilde{b}\;,\quad \alpha=-2\;,
\end{align}
{with one undetermined coefficient $\tilde{b}$.}
By applying the modified version of the algorithm in \cite{Chen:2017yze} explained in the last subsection to $c=25$, we will see that this asymptotic scaling of the formula is correct, and also that the analytically determined coefficient is correct.
{We will also numerically see that $\tilde{b}=0$, which fixes the undetermined four-derivative coefficient in the $6d$ moduli EFT.}

Let us first fit the numerical data for $\lambda_n$ to the asymptotic formula
\begin{equation}
    \begin{split}
        \log \lambda_n=\log\left(\Gamma(2n+1)\right)+ An+\upbeta n^{1/2}+\upalpha \log n +B+\upgamma n^{-1/2}+O\left(n^{-1}\right)
    \end{split}
    \label{eq:6Dexpectation}
\end{equation}
at large-$n$.
More precisely, we define the discrete second derivative of a function as $\delta^2 F_n\equiv F_{n+1}-2F_{n}+F_{n-1}$, and fit it with
\begin{align}\label{eq:tofit}
    \begin{split}
        \delta^2R_n&\equiv \delta^2 \left(\log \lambda_n- \log\Gamma(2n+1)\right)
        =\delta^2 \log \lambda_n- \log\left(\frac{(2n+2)(2n+1)}{2n(2n-1)}\right)\\
        &=-\frac{\upbeta}{4}n^{-3/2}-{\upalpha}{n^{-2}}+\frac{3}{4}\upgamma n^{-5/2}+\updelta n^{-3}+\upepsilon n^{-7/2}+O(n^{-4})
    \end{split}
\end{align}
This has the advantage of eliminating $A$ and $B$ from the formula, which are convention- and scheme-dependent in terms of the $6d$ EFT.

\begin{figure}[t!]
    \begin{subfigure}[t!]{0.5\columnwidth}
        \begin{center}
            \begin{overpic}[ width=0.97\columnwidth]{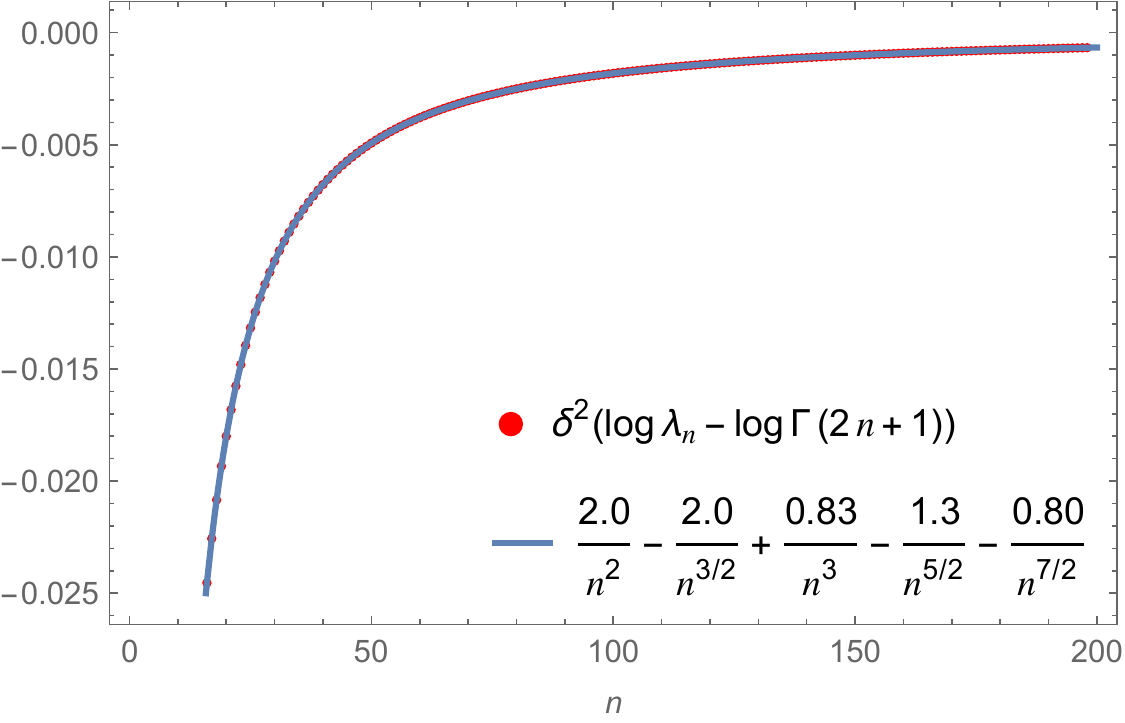}
            \end{overpic}
        \end{center}
    \end{subfigure}
    \hfill
    \begin{subfigure}[t!]{0.5\columnwidth}
        \begin{center}
            \begin{overpic}[width=0.97\columnwidth]{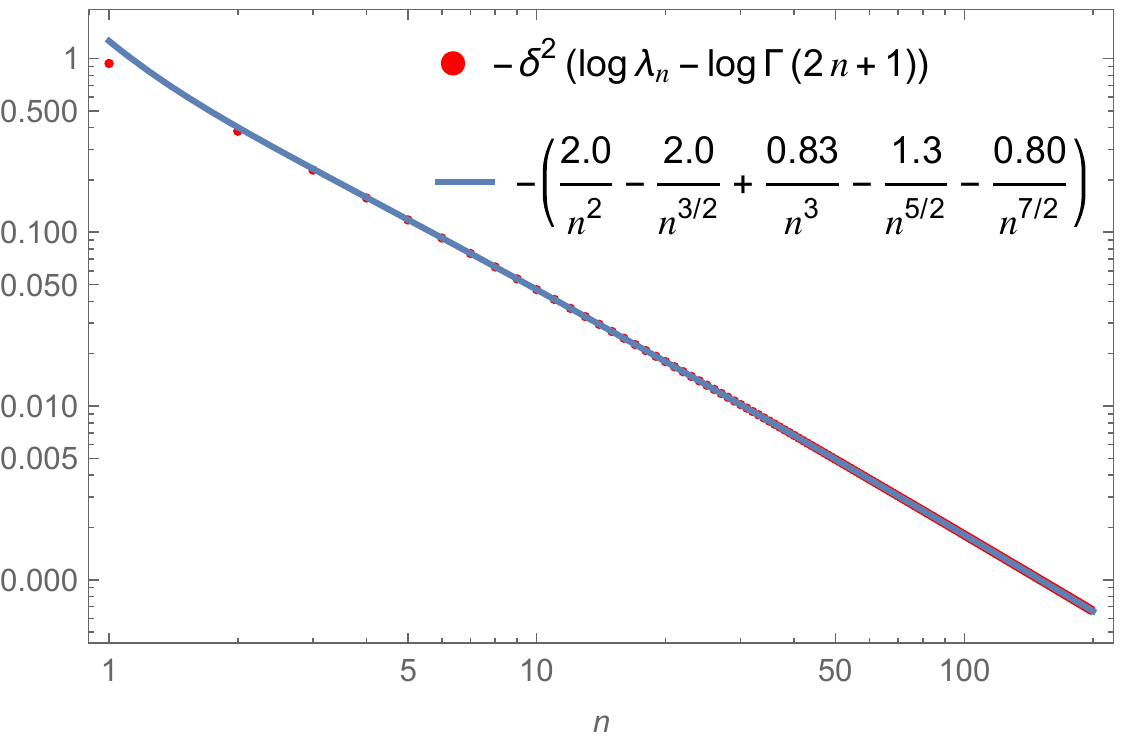}
            \end{overpic}
            \end{center}
    \end{subfigure}
    \caption{Comparison between $\delta^2 \left(\log \lambda_n- \log\Gamma(2n+1)\right)$ and the fit $-\frac{\upbeta}{4}n^{-3/2}-{\upalpha}{n^{-2}}+\frac{3}{4}\upgamma n^{-5/2}+\updelta n^{-3}+\upepsilon n^{-7/2}$, where the values are given in \eqref{tab:my_label}. The left is shown in the linear scale, whereas the right in log-log scale.}
    \label{fig:leading}
\end{figure}

Our result for the fit becomes the following,
\begin{equation}
    \begin{tabular}{c|c}
         & {Estimate}  \\
         \hline\hline
        $\upbeta$ & $7.9998$  \\
        $\upalpha$ & $-1.9996$  \\
        $\upgamma$ & $-1.7193$ \\
        $\updelta$ & $0.8333$ \\
        $\upepsilon$ & $-0.8036$
    \end{tabular}
    \label{tab:my_label}
\end{equation}
We show the result of the fit in a graph in Figure \ref{fig:leading}.
We first see that the value of $\upalpha$ is consistent with $\alpha=-2$.
Secondly, we will take the value of $\upbeta$ to indicate that $\beta=8$.
It implies $\tilde{b}=0$ from \eqref{eq:OPE} and hence we conclude that $L_{\partial^4}=b(2\mathcal{O}_2-\mathcal{O}_1)$. As anticipated above, this is the four-derivative term of a DBI-type action.

\begin{figure}[t!]
    \begin{subfigure}[t!]{0.5\columnwidth}
        \begin{center}
            \begin{overpic}[ width=0.97\columnwidth]{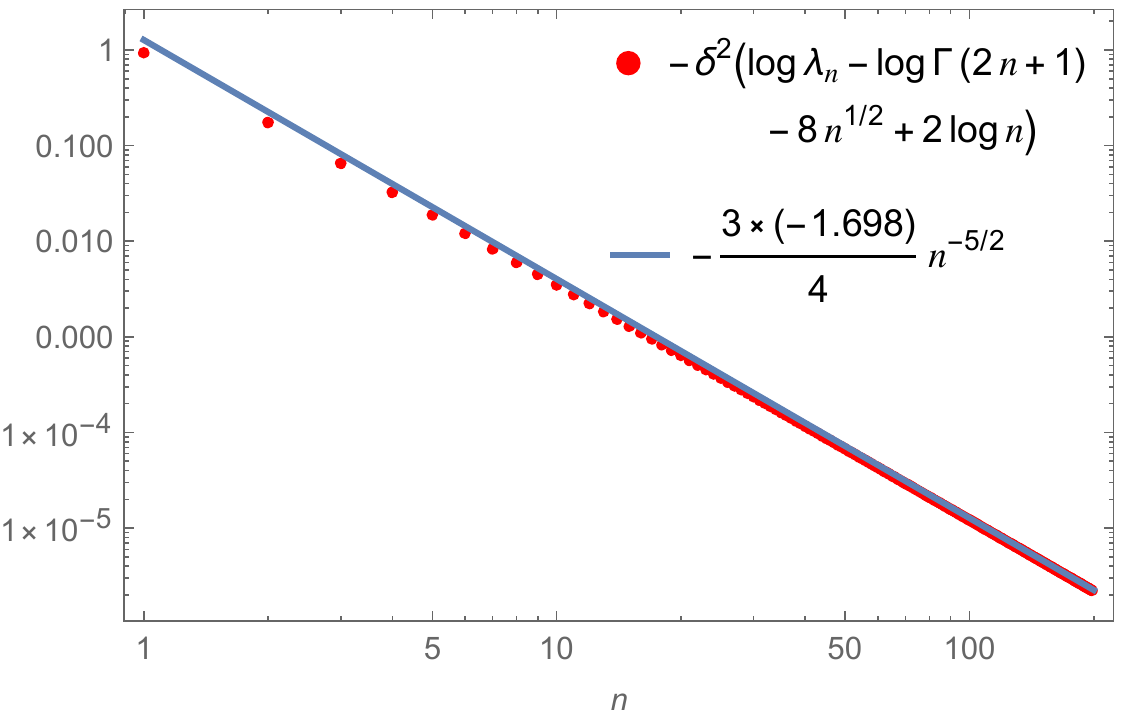}
            \end{overpic}
        \end{center}
    \end{subfigure}
    \begin{subfigure}[t!]{0.5\columnwidth}
        \begin{center}
            \begin{overpic}[width=0.97\columnwidth]{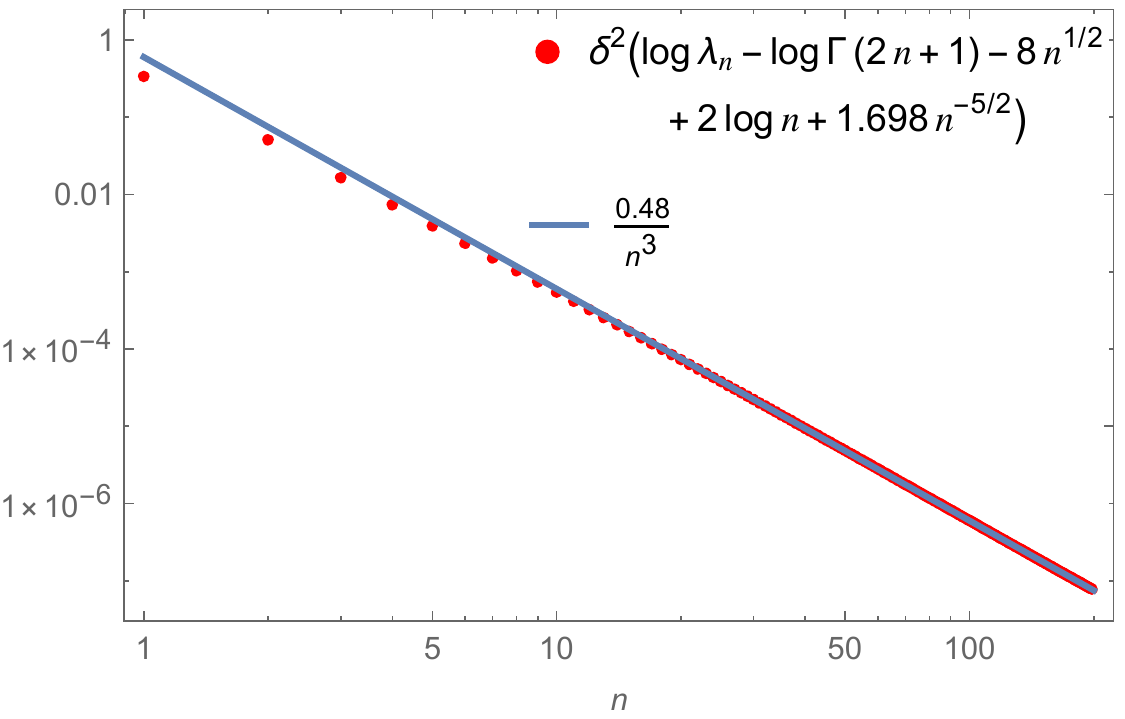}
            \end{overpic}
            \end{center}
    \end{subfigure}
    \caption{(Left) Log-log plot of $-\delta^2 R_n^{[1]}$ compared to the function scaling as $O(n^{-5/2})$. (Right) Log-log plot of $\delta^2 R_n^{[2]}$ compared to the function scaling as $O(n^{-3})$. These justify our claim that $\alpha=-2$ and $\beta=8$.}
    \label{fig:res}
\end{figure}

Let us now assume that $\beta=8$ and $\alpha=-2$.
We now plot the difference 
\begin{align}
    \begin{split}
        \delta^2R_n^{[1]}&\equiv
        \delta^2 \left(\log \lambda_n- \log\Gamma(2n+1)-\beta n^{1/2}-\alpha \log n\right)\biggr|_{\alpha=-2,\,\beta=8}\\
        &=\delta^2R_n-\left(-\frac{\beta}{4}n^{-3/2}-\alpha{n^{-2}}\right)\biggr|_{\alpha=-2,\,\beta=8}.
    \end{split}
\end{align}
in Figure \ref{fig:res}.
We see that the difference quite nicely behaves as $O(n^{-5/2})$, which backs up our conclusion that $\beta=8$ and $\alpha=-2$, \it a posteriori. \rm
We also further subtracted the $O(n^{-5/2})$ piece, such that
\begin{align}
    \begin{split}
        \delta^2R_n^{[2]}&\equiv 
        \delta^2 \left(\log \lambda_n- \log\Gamma(2n+1)-\beta n^{1/2}-\alpha \log n-\gamma n^{-5/2}\right)\biggr|_{
        \begin{subarray}{l}
            \alpha=-2,\,\beta=8,\,\\ \gamma=-1.698
        \end{subarray}}
        \\
        &=\delta^2R_n-\left(-\frac{\beta}{4}n^{-3/2}-\alpha{n^{-2}}+\frac{3}{4}\gamma n^{-5/2}\right)\biggr|_{
        \begin{subarray}{l}
            \alpha=-2,\,\beta=8,\,\\ \gamma=-1.698
        \end{subarray}},
    \end{split}
\end{align}
and we observed that it scales as $O(n^{-3})$.
The result is also shown in Figure \ref{fig:res}. 

\bigskip

To conclude, we have combined the expectation from the $6d$ EFT with $2d$ numerics to argue that the two-point function of a large $R$-charge half-BPS operator can be completely determined up to $O(\log n)$, modulo the unimportant $An+B$ piece:
\begin{align}
    \log \lambda_n=\log\Gamma(2n+1)+An+8\sqrt{n}-2\log n +B+O(n^{-1/2}).
\end{align}
In principle we can also fix the coefficients $A$ and $B$ numerically. 
We will fix $A$ in the next section.

\section{Generalization to Higher Rank and Central Charge}\label{sec:higher_rank}

\subsection{\texorpdfstring{$\lambda_n$}{lambdan} for General Central Charge \texorpdfstring{$c$}{c}}

In principle one can now move on to higher-rank $\mathcal{N}=(2,0)$ theories and attempt to compute the same OPE coefficient $\lambda_n$ using these ideas, but there are several obstacles to doing so. First, the EFT at higher rank is more complicated, with more terms allowed already at the 4-derivative level \cite{Cordova:2015vwa}. In addition, the half-BPS chiral ring is larger, and in particular there are multiple operators which have the same protected dimension at any $R$-charge. As a result, we expect multiple saddles to appear, and it is not immediately clear which saddle should correspond to which operator (see \cite{2405.19043} for progress in $4d$ theories). However, we will show that a generic large-charge analysis still provides us with enough analytic tools to fix $\lambda_n$ numerically.

For central charges $c=4r^3+12r^2+9r$ for an integer $r$, the operator $\T{n}$ again corresponds to the $1/2$-BPS operator $\Phi^n$ in the $6d$ $\mathcal{N}=(2,0)$ $A_r$ SCFT. Briefly, this is because $\Phi$ is still mapped to $\T{1}=T$, and so $\Phi^n$ should map to an operator appearing in the OPE of $n$ $T$'s. The same arguments as above then fix this operator to be $\T{n}$.\footnote{We thank L. Rastelli for discussions on this point.} This information is enough to learn some general lessons from the large charge expansion, if we assume that there exists some saddle of the moduli space EFT which corresponds to this operator. We assume that this saddle behaves schematically as above, with the scaling $\abs{\varphi}= O(\sqrt n)$ while the other moduli have VEVs of order 1. Returning to the calculation in section \ref{sec:OPE}, we note that the leading contribution $n\log n$ comes from the source term, with the kinetic term contributing a term proportional to $n$. These facts only rely on the form of our saddle and the fact that the metric is flat, which extends to higher-rank cases as well. Beyond this term dimensional analysis still predicts an expansion in $1/\sqrt n$. As a result, we obtain the following ansatz for the OPE coefficient $\lambda_n$ at higher central charge: 
\begin{equation}\label{eq:ansatz}
    \log\lambda_n=\log\Gamma(2n+1)+A n+\beta \sqrt n+\alpha\log n+B+O(n^{-1/2})\;,
\end{equation}
for some unknown coefficients $A,\,B,\,\alpha,\,\beta$.
$A$ and $B$ are normalization- and scheme-dependent as in the rank-one case. Even though this ansatz was obtained only for central charges which correspond to some $6d$ theory, we conjecture that it extends to other central charges as well.

With this ansatz in hand, we can now turn to the corresponding $2d$ calculation to check it numerically.
With details regarding the method of the fit given in Appendix \ref{app:numnum}, we report a surprising conjecture borne out by the consideration above --
Namely, our $2d$ numerics are consistent with the result
\begin{equation}\label{eq:2d_results}
    \log\lambda_n=\log \Gamma(2n+1)-2n\log 2 +4\sqrt{\frac{c-1}{12}}\sqrt{2n} -\frac{c-1}{12}\log n+B+O(n^{-1/2})\;,
\end{equation}
for any central charge $c>1$.\footnote{It is clear that the results could not be valid for all $c<1$, since \textit{e.g.},~the minimal models all have $\braket{ \T{n}\T{n} }=0$ for large enough $n$ (as discussed in Appendix \ref{app:Tn}).}

It would be very interesting to try to obtain \eqref{eq:2d_results} analytically directly in $2d$ for all $c>1$. The coefficient $\frac{c-1}{12}$ in front of the $\log$ term is highly suggestive, and the leading $\log \Gamma(2n+1)$ term seems to indicate that $\T{n}$ behave as ($2n$) free fields at high enough $n$. We leave this analysis for future work. 

\subsection{A Stringy Contribution at \texorpdfstring{$c=1$}{c=1}?}
\label{sec:worldsheet_instanton}

\begin{figure}[t!]
    \centering
    \includegraphics[width=0.7\linewidth]{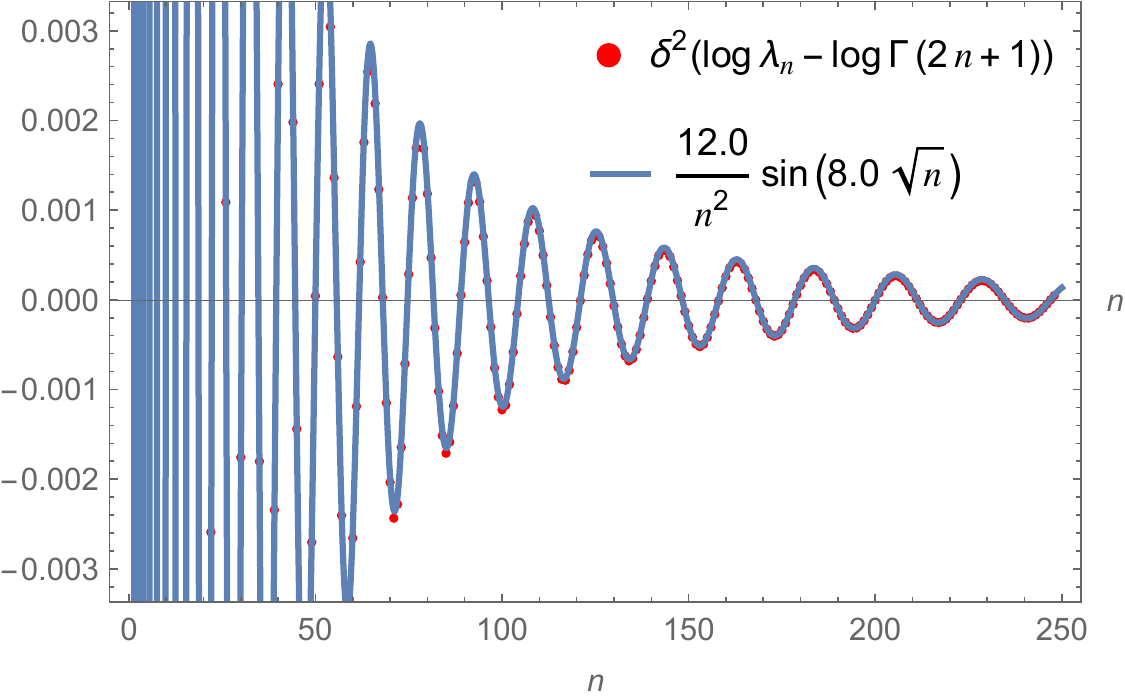}
    \caption{Comparison between $\delta^2\left(\log\lambda_n-\log \Gamma(2n+1)\right)$ and $\frac{12.0}{n^{2}}\sin\left(8.0\sqrt{n}\right)$. The oscillation might be interpreted as the imaginary-tension BPS string worldsheet instanton correction to $\log \lambda_n$ of a hypothetical non-unitary $6d$ interacting SCFT corresponding to $c=1$.}
    \label{fig:osc}
\end{figure}

We comment on a strange surprise at $c=1$. Setting $c=1$ into our result for $\lambda_n$ in \eqref{eq:2d_results} we find
\begin{equation}
    \log\lambda_n=\log \Gamma(2n+1)+2n\log 2+B+O(n^{-1/2})\;.
\end{equation}
We can now ask about subleading corrections to this expression. Numerically we find the following contributions:
\begin{equation}
    \delta^2\left(\log\lambda_n-\log \Gamma(2n+1)\right)
    =\frac{12.0}{n^2}\sin\left(8.0\sqrt{n}\right)+O(n^{-5/2}).
\end{equation}
The result of the fit is shown in Fig. \ref{fig:osc}.
Note that the actual numerical result is at $c=1.001$ because of a numerical breakdown similar to the one we experienced for $c=25$.
In terms of $\log \lambda_n$, this means that there is a contribution which goes as $\frac{0.75}{n}\sin(8.0\sqrt{n})$, so that
\begin{align}
    \log\lambda_n=\log\Gamma(2n+1)+2n\log 2+B+ \frac{0.75}{n}\sin(8.0\sqrt{n})+O(n^{-1/2})
\end{align}

Let us assume that there exists some (potentially nonunitary) interacting $6d$ SCFT which under the $6d$/$2d$ correspondence maps to the $c=1$ theory. Then this contribution is consistent with a correction of the form $\sim\frac{1}{n}e^{i\sqrt n}$, which corresponds to a complex instanton correction. We identify it as a worldsheet instanton since in our EFT $\braket{ \phi }\sim \sqrt n$, and since $\phi$ has scaling dimension 2 we identify $\sqrt n$ with the worldsheet area. 
The additional oscillating term may thus be interpreted as coming from a worldsheet instanton in a hypothetical non-unitary $6d$ interacting SCFT, in analogy with the worldline instanton correction for $4d$ $\mathcal{N}=2$ SCFTs \cite{1804.01535,1908.10306,2005.03021,2103.05642,2103.09312}. 
While it is not clear if we can identify such a contribution as a genuine stringy correction in $6d$ when we set $c=1$, presumably such a correction appears also at higher $c$, and in particular appears at $c=25$, where it might correspond to a physical stringy correction scaling as $O(e^{-\sqrt{n}})$. However, understanding these subleading terms becomes difficult numerically as we increase $c$, and so we cannot identify this term without more analytic work to fix other subleading terms analytically first. We leave this analysis to future work.

\section{Conclusions and Outlook}\label{sec:future}

In this paper we initiated the study of $6d$ SCFTs at large charge.
We studied observables using standard large-charge methods in combination with methods from the $6d$/$2d$ correspondence, to determine the two-point function of large $R$-charge half-BPS operators $\Phi^n$ of the $A_1$ theory, $\lambda_n\equiv \abs{x}^{8n}\braket{\Phi^n(x)\Phi^n(0)}$.
As our argument was complicated, involving $6d$ analytic result with $2d$ numerics, we summarize it below.

We first computed $\lambda_n$ from the moduli space effective action of the $A_1$ theory.
This is in principle completely determined by $\mathcal{N}=(2,0)$ SUSY up to six-derivatives without undetermined parameters.
{However, we are not able to completely fix the form of the four-derivative effective operator (even though this should be possible with much but finite effort) and unfortunately left one undetermined coefficient $\tilde{b}$.}
The EFT can then be used to compute $\lambda_n$ as the partition function on $S^6$ with insertions of large $R$-charge operators at antipodal points.
After some computations, we found
\begin{align}\label{eq:lambda_conclusion}
    \log \lambda_n
    &=\log \Gamma(2n+1) + An+\left(8-160\pi^{3/2}\tilde{b}\right)n^{1/2}-2\log n+B+O(n^{-1/2})\;.
\end{align}

We then used numerics in $2d$ by appealing to the correspondence between the half-BPS ring of $6d$ $\mathcal{N}=(2,0)$ SCFTs and $2d$ chiral algebras. In particular, it is known that the $A_1$ theory corresponds to a Virasoro algebra with central charge $c=25$.
It was crucial to notice that the $2n$-th Taylor expansion coefficient of the vacuum Virasoro block contains a piece proportional to $h^{2n}$, and the coefficient of this was inversely related to $\lambda_n$.
We used the fast numerical algorithm of \cite{Chen:2017yze} and computed the expansion coefficients of the Virasoro block at $c=25$.
{We then found that $\tilde{b}=0$, and so we determined that}
\begin{align}
    \log \lambda_n=\log\Gamma(2n+1)+An+8\sqrt{n}-2\log n +B+O(n^{-1/2}).
\end{align}

We have also generalized our formula to higher-rank interacting $6d$ SCFTs.
As they have different central charges $c$ in terms of the $6d/2d$ correspondence, we simply numerically computed $\lambda_n$ for various $c$.
We numerically found that 
\begin{equation}\label{eq:conc_res}
    \log\lambda_n=\log \Gamma(2n+1)-2n\log 2 +4\sqrt{\frac{c-1}{12}}\sqrt{2n} -\frac{c-1}{12}\log n+B+O(n^{-1/2})\;,
\end{equation}
which we left as a conjecture to be proven analytically from both $6d$ and from $2d$.

We further identified a BPS worldsheet instanton correction to the formula at $c=1$.
The contribution to $\log \lambda_n$ at $c=1$ numerically turned out to be $\frac{0.75}{n}\sin\left(8.0\sqrt{n}\right)$.
We argued that it can be interpreted in $6d$ as a contribution from the imaginary-tension BPS worldsheet instanton, which is present in a fictitious $6d$ $\mathcal{N}=(2,0)$ non-unitary but interacting SCFT.

We list some open questions and future directions:
\begin{enumerate}
    \item It would be nice to derive our main result \eqref{eq:conc_res} directly in $2d$.
    \item As discussed above, the spectrum of non-protected operators is very mysterious for these $6d$ theories. The large-charge EFT we have derived can be used to find the dimension of the lowest-dimension unprotected operator at large charge. This computation is slightly involved since operators just above the BPS bound are still protected and we require $\Delta > \Delta_{\text{BPS}}+6$ to find an unprotected operator \cite{Cordova:2016emh}. 
    \item The EFT we have derived is in principle enough in order to derive $\lambda_n$ to all orders in $1/n$, since all additional terms in the EFT are $D$-terms which do not contribute to $\lambda_n$. In practice this requires resumming all quantum corrections, which is difficult. For $4d$ theories, a recursion relation for correlators was used in \cite{1804.01535,1908.10306,2005.03021,2103.05642,2103.09312} to perform this computation and resum all such corrections, giving an all-orders answer. Such a recursion relation is not yet known for $6d$ theories, and so it is not clear whether an analogous computation can be performed. 
    \item A different limit that is well-studied for $6d$ $\mathcal{N}=(2,0)$ SCFTs is the holographic limit of large rank (corresponding to large central charge in $6d$), and it would be interesting to compare results for these two limits and find how one can interpolate between them.
    \item Further studies of the subleading worldsheet-instanton-like corrections to our large-charge formula discussed in section \ref{sec:worldsheet_instanton} would be very interesting from the point of view of both $6d$ and $2d$. In terms of $6d$, it would be in principle possible to extract such terms by considering the action of the BPS string wrapping $S^2$ inside spacetime $S^6$ -- even though such a straightforward computation could be very tedious.
    In terms of $2d$, it would also be possible to determine the (complex) tension of the string from numerics in terms of $c$.
    One could then hope to analytically continue the tension to $c=25$, which corresponds to the physical $A_1$ theory.
    \item A similar large-charge analysis can be done for Higgs branches in $4d$ $\mathcal{N}=2$ theories. While Coulomb branches have been discussed extensively, a study of Higgs branches is still lacking. One can compare with the expectations from the $4d$/$2d$ correspondence \cite{Beem:2013sza}, in analogy with the analysis done in this paper. 
\end{enumerate}

\section*{Acknowledgements}
The authors would like to thank Fernando Alday, Gabriel Cuomo, Simeon Hellerman, and Yuya Kusuki for useful conversations. The authors are especially grateful to Arash A. Ardehali and Leonardo Rastelli for many helpful discussions and for sharing unpublished work. 
The authors also thank the ``21st Simons Physics Summer Workshop'' where this work was completed. MW thanks the hospitality of Simons Center for Geometry and Physics while part of this work was in progress. The work of JJH is supported by DOE (HEP) Award DE-SC0013528 and BSF grant 2022100. The work of MW is supported by Grant-in-Aid for JSPS Fellows (No.~22KJ1777) and by MEXT KAKENHI Grant (No.~24H00957).

\appendix

\section{Determination of \texorpdfstring{$\mathcal{O}_{1,2}$}{O1, O2}}
\label{sec:o12}

We determine the form of $\mathcal{O}_{1,2}$.
At four-derivative level all operators must take one of the following forms (up to the leading order EOM and total derivatives),
\begin{align}
    &f_{IJKL}(\varphi)\partial_\mu \varphi^I \partial^\mu \varphi^J \partial_\nu \varphi^K \partial^\nu \varphi^L \label{eq:f}\\
    &g_{IJK}(\varphi)\partial_\mu \varphi^I \partial_\nu \varphi^J \partial^\mu\partial^\nu \varphi^K \label{eq:g}\\
    &h_{IJ}(\varphi)\partial_\mu\partial_\nu \varphi^I\partial^\mu\partial^\nu \varphi^J \label{eq:h}
\end{align}
where $f$, $g$, and $h$ are some functions.

SUSY constrains the form of $f_{IJKL}$ in \eqref{eq:f} as well as excludes the possibilities for \eqref{eq:g} and \eqref{eq:h}, as argued in \cite{Cordova:2015vwa} and as we now briefly review.
By expanding $\varphi^I$ around a homogeneous VEV such that $\varphi^I=\braket{\varphi^I}+\varphi_f^I$, each effective operator is expanded in terms of the number of fluctuations, all of which need to respect $\mathcal{N}=(2,0)$ SUSY again.
Now, the allowed four-derivative SUSY-preserving deformations constructed of a single Abelian tensor multiplet contains $n\geq 4$ fields and must transform as a traceless symmetric $(n-4)$-tensor of $SO(5)_R$.
This rules out the terms \eqref{eq:g} and \eqref{eq:h} at leading order in the fluctuation expansion because they contain less than four fields:
\begin{align}
    &g_{IJK}(\varphi)\partial_\mu \varphi^I \partial_\nu \varphi^J \partial^\mu\partial^\nu \varphi^K
    \ni g_{IJK}(\braket{\varphi})\partial_\mu \varphi^I_f \partial_\nu \varphi^J_f \partial^\mu\partial^\nu \varphi^K_f\\
    &h_{IJ}(\varphi)\partial_\mu\partial_\nu \varphi^I\partial^\mu\partial^\nu \varphi^J
    \ni h_{IJ}(\braket{\varphi})\partial_\mu\partial_\nu \varphi^I_f\partial^\mu\partial^\nu \varphi^J_f
\end{align}
and so are not allowed as SUSY-preserving deformations.

It also constrains the form of $f_{IJKL}$.
At leading order in the fluctuation expansion, \eqref{eq:f} needs to produce an $SO(5)_R$-invariant deformation (\it i.e., \rm a zero-tensor) with four fields, so that $f_{IJKL}(\varphi)$ needs to be proportional to either $\delta_{IJ}\delta_{KL}$ or $\delta_{IK}\delta_{JL}$.
The overall function multiplying the Kronecker deltas are determined by conformal symmetry and $R$-charge neutrality to be $1/\abs{\varphi}^3$. 
We therefore obtain two candidate operators,
\begin{align}
    \mathcal{O}_{1}&\equiv \frac{\partial_\mu \varphi^I \partial^\mu \varphi^I \partial_\nu \varphi^J \partial^\nu \varphi^J}{\abs{\varphi}^3}\sim \frac{\left(\partial^2 \abs{\varphi}^2\right)^2}{4\abs{\varphi}^3}\\
    \mathcal{O}_{2}&\equiv \frac{\partial_\mu \varphi^I \partial^\mu \varphi^J \partial_\nu \varphi^I \partial^\nu \varphi^J}{\abs{\varphi}^3}\sim \frac{\partial^2 (\varphi^I\varphi^J) \partial^2 (\varphi^I\varphi^J)}{4\abs{\varphi}^3},
\end{align}
where $\sim$ refers to an equality modulo the leading order EOM and total derivatives.

\section{Evaluation of  \texorpdfstring{$S_{4,6}$}{S4, S6} on the Saddle-Point}

\subsection{Evaluation of \texorpdfstring{$S_4$}{S4}}
\label{app:four-derivative}

Let us now evaluate
{
\begin{align}
    S_4\equiv b_1S_{4,1}+b_2S_{4,2}\equiv b_1\int d^6 x
    \frac{\left(\partial^2 \abs{\varphi}^2\right)^2}{4\abs{\varphi}^3}
    +b_2\int d^6 x
    \frac{\partial^2 (\varphi^I\varphi^J) \partial^2 (\varphi^I\varphi^J)}{4\abs{\varphi}^3}
\end{align}
}
on the leading order saddle-point \eqref{eq:6d_sol} which we repeat here
\begin{equation}
    \phi(x)=\frac{e^{i \beta_0}x_{12}^2}{\sqrt{4\pi^3}(x-x_2)^4} (2n)^{1 / 2}, \quad \phibar(x)=\frac{e^{-i \beta_0}x_{12}^2}{\sqrt{4\pi^3}(x-x_1)^4} (2n)^{1 / 2}\;.  
\end{equation}

\paragraph{Evaluation of $S_{4,1}$}
First of all, let us rewrite $S_{4,1}$ by using the leading order free EOM,
\begin{align}
    S_{4,1}\equiv \int d^6 x\frac{\left(\partial^2 \abs{\varphi}^2\right)^2}{4\abs{\varphi}^3}
    \sim 
    \int d^6 x\frac{\left(\partial_\mu {\varphi^I}\partial^\mu{\varphi^I}\right)^2}{\abs{\varphi}^3}\;.
\end{align}
Truncating the part containing $\varphi^{3,4,5}$, we have
\begin{align}
    \begin{split}
        S_{4,1}\biggr|_{\rm saddle}&=\int d^6 x\frac{\sqrt{2}\left(\partial_\mu {\phi}\partial^\mu{\phibar}\right)^2}{\abs{\phi}^3}\biggr|_{\rm saddle}\\
        &=
        4^4\pi^{-3/2}n^{1/2}x_{12}^2\int d^6 x\frac{\left((x-x_1)_\mu(x-x_2)^\mu\right)^2}{\abs{x-x_1}^6\abs{x-x_2}^6}\;.
    \end{split}
\end{align}
where $x_{12}\equiv \abs{x_1-x_2}$.

From now on, let us set $x_1=(0,0,0,0,0,1)$ and $x_2=0$ without loss of generality because of conformal invariance.
We can then take the spherical coordinates,
\begin{align}
    \begin{split}
        S_{4,1}\biggr|_{\rm saddle}&=
        4^4\pi^{-3/2}n^{1/2}\int d^6 x\frac{\left((x-x_1)_\mu x^\mu\right)^2}{\abs{x-x_1}^6\abs{x}^6}\\
        &=4^4\pi^{-3/2}\mathop{\tt Vol}(S^4)n^{1/2}
        \int r^5 \sin^4 \theta dr d\theta\, 
        \frac{(r^2-r\cos\theta)^2}{r^6\left(r^2 + 1 - 2r\cos\theta \right)^3}\;,
    \end{split}
\end{align}
where in the second line we have already integrated over the homogeneous $S^4$ direction and the range of integration is $0\leq r<\infty$ and $0\leq\theta\leq \pi$.
This can be evaluated as
{
\begin{align}\label{eq:S41_evaluated}
    \begin{split}
        S_{4,1}\biggr|_{\rm saddle}&=4^4\pi^{-3/2}\mathop{\tt Vol}(S^4)\sqrt{n}
        \int r^5 \sin^4 \theta dr d\theta\, 
        \frac{(r^2-r\cos\theta)^2}{r^6\left(r^2 + 1 - 2r\cos\theta \right)^3}\\
        &=
        4^4\pi^{-3/2}\mathop{\tt Vol}(S^4)\sqrt{n}
        \int r^5 dr \, F(r)=96\pi^{3/2}\sqrt{n}\;,
    \end{split}
\end{align}
}
where
{
\begin{align}\label{eq:angular_integral}
    F(r)=
    \begin{cases}
        \frac{\pi}{16r^4} & 0<r<1\\
        \frac{\pi(6r^2-5)}{16r^{10}} & r>1
    \end{cases}
\end{align}

}
\paragraph{Evaluation of $S_{4,2}$}

Let us first simplify $S_{4,2}$ by truncating the part containing $\varphi^{3,4,5}$ to be
\begin{align}\label{eq:S42_reduction}
    S_{4,2}\biggr|_{\rm saddle}=\int d^6 x\frac{\sqrt{2}}{8\abs{\phi}^3}\left[\partial^2(\phi^2)\partial^2(\phibar^2)+\left(\partial^2(\phi\phibar)\right)^2\right]\biggr|_{\rm saddle}\;.
\end{align}
This can be further simplified upon using the EOM for $\phi$ as
\begin{align}\label{eq:S42_decomposition}
    S_{4,2}\biggr|_{\rm saddle}=\frac{1}{2}S_{4,1}\biggr|_{\rm saddle}+8\sqrt{2}\int d^6 x\,\sqrt{\phibar}\partial^2\partial^2\sqrt{\phi}\biggr|_{\rm saddle}\;.
\end{align}
{Now, without loss of generality we set $x_2=0$ and take $\abs{x_1}\to\infty$.}
{One can then ignore everything which is subleading in $1/\abs{x_1}$ in the second integral, such that}
\begin{align}\label{eq:S42_evaluated}
    \begin{split}
        S_{4,2}\biggr|_{\rm saddle}&=\frac{1}{2}S_{4,1}\biggr|_{\rm saddle}+8\sqrt{2}\int d^6 x\,\sqrt{\phibar}\partial^2\partial^2\sqrt{\phi}\biggr|_{\rm saddle}\\
        &=48\pi^{3/2}\sqrt{n}+8\pi^{-3/2}\sqrt{n}\int d^6 x\,16\pi^3\delta(x)=176\pi^{3/2}\sqrt{n}\;.
    \end{split}
\end{align}

\bigskip

To conclude, we get
\begin{align}\label{eq:S4_saddle_appendix}
    S_4\biggr|_{\rm saddle}=(96b_1+176b_2)\pi^{3/2}\sqrt {n}\;.
\end{align}

\subsection{Evaluation of \texorpdfstring{$S_6$}{S6}}
\label{app:six-derivative}

We evaluate
\begin{align}
    S_6=\Delta a\int d^6 x\, \uptau E_6
\end{align}
on the saddle-point, on $S^6$.
The relevant definitions and formulae are
\begin{align}
    \begin{split}
        \uptau\equiv -\frac{1}{2}\log \abs{\varphi}, \quad \abs{\varphi}\biggr|_{\rm saddle}=O(\sqrt{n})\\
        E_6\biggr|_{S^6}=720, \quad \mathop{\tt Vol}(S^6)=\frac{16\pi^3}{15}\\
        \Delta a = a_1-a_{U(1)}=\frac{96}{7}a_{U(1)}=\frac{2}{3}\frac{1}{(4\pi)^3}
    \end{split}
\end{align}
from which we get
\begin{align}
    S_6\biggr|_{\rm saddle} =-2\log n +O(1)\;.
\end{align}

\section{More About \texorpdfstring{$\T{n}$}{Tn}}\label{app:Tn}

We discuss various facts about $\T{n}$ and its two-point function. 

\subsection{Explicit Form of \texorpdfstring{$\T{n}$}{Tn}}

We can compute $\T{n}$ explicitly for small values of $n$. We present some results here.
\begin{align}
    \T{1}&=T\;,\\
    \T{2}&=T^2-\frac{3}{10}T''\;,\\
    \T{3}&=T^3+\frac{93}{29+70 c} T^{\prime 2}-\frac{3(67+42 c)}{2(29+70 c)} T^{\prime \prime} T-\frac{13+10 c}{4(29+70 c)} T^{\prime \prime \prime \prime}.\\
    \begin{split}
        \T{4}&=T^4
        +\frac{12 (465 c-127)}{1050 c^2+3305 c-251}(T')^2T
        -\frac{226800 c^2+1249560 c-200520}{120 \left(1050 c^2+3305 c-251\right)}T''T^2\\
        &\quad -\frac{-34020 c^2-214650 c+135270}{120 \left(1050 c^2+3305 c-251\right)}(T'')^2
        -\frac{30240 c-116880}{120 \left(1050 c^2+3305 c-251\right)}T'''T'\\
        &\quad -\frac{18000 c^2+68280 c+8040}{120 \left(1050 c^2+3305 c-251\right)}T''''T
        -\frac{630 c^2+2315 c-543}{120 \left(1050 c^2+3305 c-251\right)}T''''''.
    \end{split}
\end{align}

\subsection{Computing \texorpdfstring{$\braket{ \T{n}\T{n} }$}{<TnTn>}}

There are many ways to compute $\lambda_n=\braket{ \T{n}(1)\T{n}(0) }$ beyond the method discussed in the main text. These are less efficient than the method we use, but are more natural at small $n$ and also shed light on some physical properties of $\T{n}$. We list some of the methods we found useful:
\begin{enumerate}
    \item The first way is the ``direct'' way of computing the two-point function. We compute the OPE of two $\T{n}(x)\T{n}(0)$ and extract the leading nonsingular term, proportional to $x^0$. This can be done by hand efficiently \textit{e.g.},~by using Thielemans' OPEdefs mathematica package (see section 3.4 of \cite{Thielemans:1994er}). 
    \item One can also bypass finding the explicit form of $\T{n}$, and instead it is enough to find the general form for a quasi-primary and then use the Gram-Schmidt algorithm to extract $\lambda_n$.
    \item Another way is described in Appendix C of \cite{1710.10458}. Instead of writing $\T{n}=T^n+...$, we instead write it in terms of Virasoro generators $\T{n}=L_{-2}^n+...$ and then use their explicit commutation relations to compute the two-point function.
    \item Finally, we can relate $\lambda_n$ to the inverse of the Kac matrix, leading to a more efficient algorithm than the above (although much less efficient than the algorithm discussed in the main text). Compute the Kac matrix in the vacuum
    \begin{equation}
        M_{\{k\},\{k;\}}=\braket{ L_{\{k\}}L_{-\{k'\}}}
    \end{equation}
    where $L_{k}=L_{k_1}...L_{k_n}$. Then \cite{1710.10458} proves that (see equation 124)
    \begin{equation}
        \frac{1}{\lambda_n^2}\braket{ \T{n}\T{n} }=[M^{-1}]^{\{2...2\}\{k\}}M_{\{k\},\{k'\}}[M^{-1}]^{\{2...2\}\{k'\}}=[M^{-1}]^{\{2...2\}\{2...2\}}\;,
    \end{equation}
    and so we find
    \begin{equation}
        \lambda_{n}=\frac{1}{[M^{-1}]^{\{2...2\}\{2...2\}}}\;.
    \end{equation}
\end{enumerate}

We present explicit results for $\lambda_n$ for low values of $n$ in table \ref{tab:TnTn}.
\begin{table}[t!]
\begin{center}
\begin{tabular}{ c c }
$n$ & $\lambda_n$ \\ 
\hline
1  & $\frac{c}{2}$   \\
2  & $\frac{1}{10} c (5 c+22)$   \\
3  & $\frac{3 c (2 c-1) (5 c+22) (7 c+68)}{4 (70 c+29)}$\\
4 & $\frac{3 c (2 c-1) (3 c+46) (5 c+3) (5 c+22) (7 c+68)}{2 \left(1050 c^2+3305 c-251\right)}$\\
5  &$ \frac{15 c (2 c-1) (3 c+46) (5 c+3) (5 c+22) (7 c+68) (11 c+232)}{4 \left(11550 c^2+76675 c+3767\right)}$ \\
6  & $\frac{45 c (2 c-1) (3 c+46) (5 c+3) (5 c+22) (7 c+25) (7 c+68) (10 c-7) (11 c+232) (13 c+350)}{4 \left(10510500 c^4+151698400 c^3+424911495 c^2-265783494 c+19445435\right)}$\\
7 &  $\frac{63 c (2 c-1) (3 c+46) (4 c+21) (5 c+3) (5 c+22) (5 c+164) (7 c+25) (7 c+68) (10 c-7) (11 c+232) (13 c+350)}{8 \left(42042000 c^5+1096582900 c^4+8440143100 c^3+18434067003 c^2-4906238290 c-2669202025\right)}$\\
8 &  $\frac{315 c (2 c-1) (3 c+46) (4 c+21) (5 c+3) (5 c+22) (5 c+164) (7 c+25) (7 c+68) (10 c-7) (11 c+232) (13 c+350) (17 c+658)}{2 \left(3573570000 c^5+121461770500 c^4+1222244032300 c^3+3521676058455 c^2+486995135366 c-1410211568605\right)}$
\end{tabular}
\end{center}
\caption{$\lambda_n$ for low values of $n$.}
\label{tab:TnTn}
\end{table}
Interestingly, some patterns emerge which are simple to understand (although we could not guess the full result for general $n$).\footnote{We thank Arash Ardehali for interesting discussions on this point.} The numerator of $\lambda_n$ is given by
\begin{equation}
    \prod_{q,p}\left(c-c_{q,p}\right)
\end{equation}
where $c_{q,p}$ is the central charge of the $q,p$ minimal model and we are multiplying over $q$, $p$ which are coprime and such that $(q-1)(p-1)\leq n$. Indeed, $\lambda_n$ must vanish when $c=c_{q,\,p}$ for such $q,\,p$ since in the corresponding minimal model, $\T{n}$ is a null operator. Another part that can be easy to understand is the coefficient of $c^n$ in $\lambda_n$ (\it i.e. \rm the highest power of $c$ in the $c\to\infty$ expansion), which is given by $n!/2^n$. This can be shown by computing the two-point function $\braket{ T^n T^n}$, where here by $T^n$ we mean the standard normally-ordered product of energy-momentum tensors. This can be computed \textit{e.g.}, for a free field and the result can be immediately extrapolated to general central charges.

\subsection{Relation to Vacuum Virasoro Block Coefficients}
\label{sec:vir}

We relate $\lambda_n$ to the expansion coefficients of the vacuum Virasoro block.
Let us first give the holomorphic vacuum Virasoro block expanded in terms of the cross-ratio, $z$, (see \textit{e.g.},~\cite{1502.07742})
\begin{align}
    \mathcal{F}(c,h,z)=\sum_{n=0}^{\infty} \chi_{2n}^{\rm vac}(c,h)z^{2n} {}_2F_{1}(n,n,2n;z),
\end{align}
where 
\begin{align}
    \chi_{2n}^{\rm vac}(c,h)=\sum_{\mathcal{X}\in \mathcal{M}(c,{\rm vac})}\frac{(\tilde{C}_{hh\mathcal{X}})^2}{\mathcal{N}_{\mathcal{X}}}.
\end{align}
Here, $\mathcal{X}$ is a quasi-primary operator running over the vacuum Verma module $\mathcal{M}(c,{\rm vac})$, $\mathcal{N}_{\mathcal{X}}$ is the normalization of the two-point function of $\mathcal{X}$, while $\tilde{C}_{hh\mathcal{X}}$ is the normalized OPE coefficient $\tilde{C}_{hh\mathcal{X}}=C_{hh\mathcal{X}}/\mathcal{N}_{\mathcal{X}}$.

The character $\chi_{2n}^{\rm vac}(c,h)$ has a nice expansion in terms of $1/h$.
To see this, notice that the OPE coefficient obeys
$\tilde{C}_{hhT^n}=h^n+O(h^{n-1})$ at large $h.$ 
Since by definition $\T{n}$ is the only quasi-primary in our basis of operators of $\mathcal{M}(c,{\rm vac})$ which includes $T^n$ (and all other basis elements only include lower powers of $T$), and since it is orthogonal to all other quasi-primaries, it follows that the leading $O(h^{2n})$ piece in $(\tilde{C}_{hh\mathcal{X}})^2$ is obtained from the term with $\mathcal{X}=\T{n}$.
We therefore have
\begin{align}
    \chi_{2n}^{\rm vac}(c,h)=\frac{h^{2n}}{\mathcal{N}_{\T{n}}}\left(1+O(h^{-1})\right)=\frac{h^{2n}}{\lambda_n}\left(1+O(h^{-1})\right),
\end{align}
because $\lambda_n=\mathcal{N}_{\mathbb{T}^n}$ by definition.

All this can be cast in a different way in terms of $\mathcal{F}(c,h,z)$ using the double-scaling limit where $h\to\infty$ and $z\to 0$ while fixing $zh$.
As ${}_2F_1(n,n,2n;z)=1+nz/2+O(z^2)$, we have
\begin{align}
    \mathcal{F}(c,h,z)=\sum_{n=0}^{\infty} \frac{(zh)^{2n}}{\lambda_n}\times \left(1+O(h^{-1})\right)\left(1+O(z)\right),
\end{align}
so in other words, $\lambda_n$ is related to the expansion coefficient of the Virasoro block in the double-scaling limit $h\to\infty$ and $z\to 0$ where $zh$ is fixed.

It is advantageous to rephrase this in terms of the expansion coefficients of Zamolodchikov's $H$-function, for which \cite{Chen:2017yze} provides us with a fast numerical algorithm.
The $H$-function is defined by the relation
\begin{align}
    F(c,h_i,h_p,z) \equiv (16q)^{h_p-{\frac{c-1}{24}}}z^{{\frac{c-1}{24}}}(1-z)^{{\frac{c-1}{24}} -h_2-h_3}\theta_3(q)^{{\frac{c-1}{2}}-4\sum_i h_i}H(c,h_i,h_p,q)
\end{align}
where
\begin{align}
    q = e^{-\pi {\frac{K(1-z)}{K(z)}}}, \quad K(z)=\frac{\pi}{2} {}_2F_{1}\left(\frac{1}{2},\frac{1}{2},1;z\right), \quad \theta_3(q) = \sum_{n=-\infty}^{\infty}q^{n^2}
\end{align}
while $h_{i=1,2,3,4}$ are the external dimensions and $h_p$, the internal.
For the case of our interest, we simply set $h_p=0$ and $h\equiv h_{1,2,3,4}$.
We hereafter denote the $H$-function for such a special case as $H(c,h,q)$.
Now, because
\begin{align}
    (16q)^{h_p-{\frac{c-1}{24}}}z^{{\frac{c-1}{24}}}(1-z)^{{\frac{c-1}{24}} -h_2-h_3}\theta_3(q)^{{\frac{c-1}{2}}-4\sum_i h_i}=1+O(h^{-1})
\end{align}
in the same double-scaling limit, we now have
\begin{align}
\label{eq:vir}
    H(c,h,q)=\mathcal{F}(c,h,z)\times \left(1+O(h^{-1})\right)=\sum_{n=0}^{\infty} \frac{(zh)^{2n}}{\lambda_n}\times \left(1+O(h^{-1})\right)
\end{align}
in this limit as well.
By noting that $q=z/16+O(z^2)$, we finally have
\begin{align}
    H(c,h,q)=\sum_{n=0}^{\infty} \frac{(16qh)^{2n}}{\lambda_n}\times \left(1+O(h^{-1})\right).
\end{align}

\section{Details of the Fit Leading to Conjecture \texorpdfstring{\eqref{eq:2d_results}}{}}
\label{app:numnum}

\begin{figure}[t!]
    \begin{subfigure}[t!]{0.5\columnwidth}
        \begin{center}
            \begin{overpic}[ width=0.97\columnwidth]{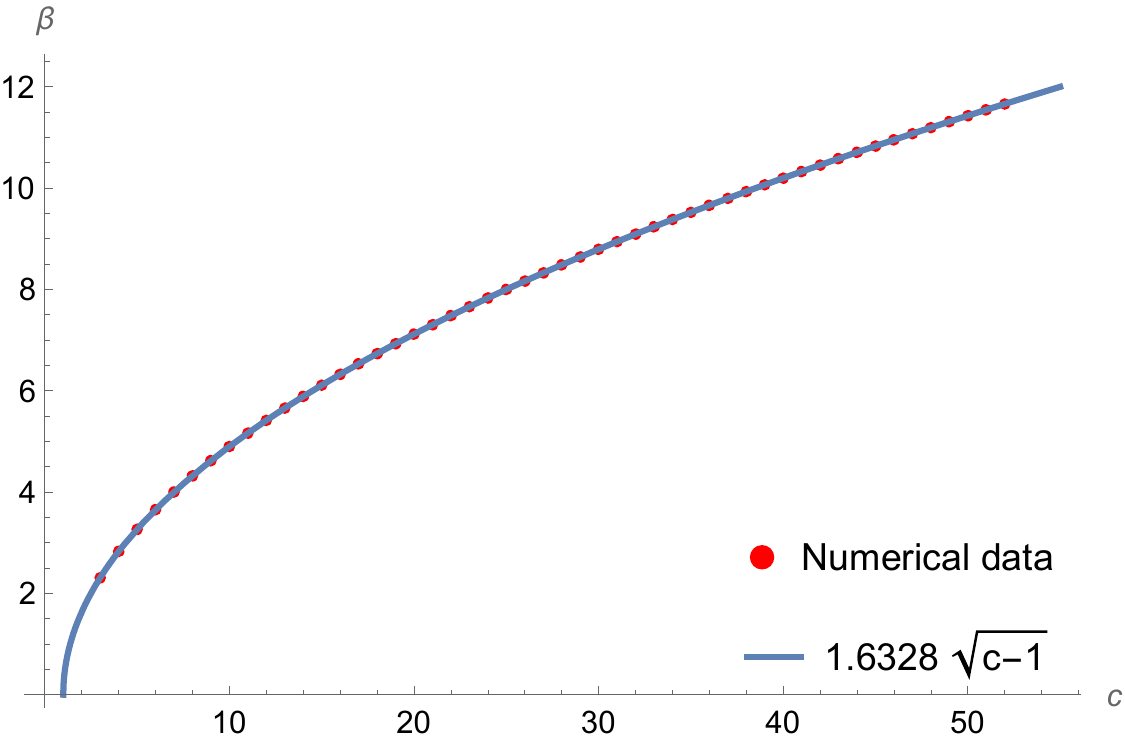}
            \end{overpic}
        \end{center}
    \end{subfigure}
    \begin{subfigure}[t!]{0.5\columnwidth}
        \begin{center}
            \begin{overpic}[width=0.97\columnwidth]{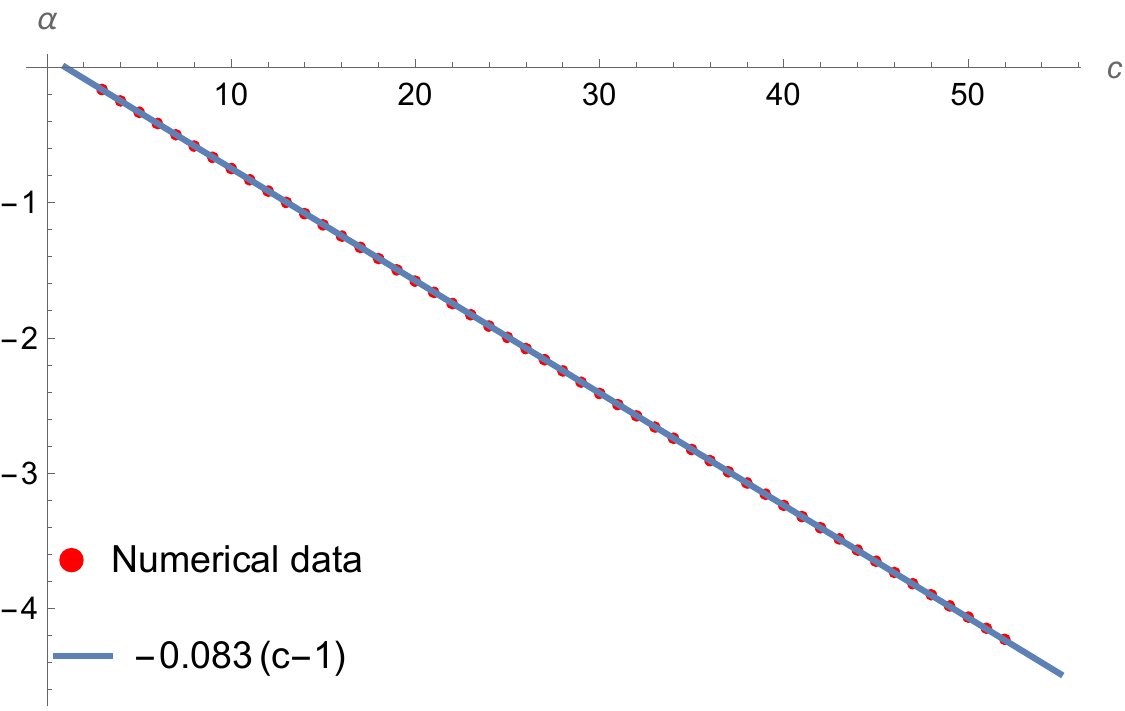}
            \end{overpic}
            \end{center}
    \end{subfigure}
    \caption{
    (Left) Numerical estimates of $\beta$ for various values of $c$ compared to $\sqrt{\frac{c-1}{12}}\times 4\sqrt{2}$.
    (Right) Numerical estimates of $\alpha$ for various values of $c$ compared to $-\frac{c-1}{12}$.
    }
    \label{fig:ccc}
\end{figure}

We explain the exact methodology of the numerics leading to \eqref{eq:2d_results}.
We first fit the residual $\delta^2 R_n$ by using
\begin{align}
    -\frac{\upbeta}{4}n^{-3/2}-{\upalpha}{n^{-2}}+\frac{3}{4}\upgamma n^{-5/2}+\updelta n^{-3}\;.
\end{align}
It turns out that $\upbeta\propto \sqrt{c-1}$ and $\upalpha\propto {c-1}$ (without any corrections, at least numerically!) with some numerical coefficients (See Figure \ref{fig:ccc}).

Using the fact that we know $\alpha=-2$ and $\beta=8$ at $c=25$, we can determine its numerical coefficients, which results in
\begin{align}
    \alpha=-\frac{c-1}{12},\quad \beta=\sqrt{\frac{c-1}{12}}\times 4\sqrt{2}\;.
\end{align}
It is very suggestive that the coefficient of the $\sqrt n$ term and the coefficient of the $\log n$ term are proportional to $\sqrt{c-1}$ and $c-1$, and in particular the fact that $\beta\propto \sqrt{\alpha}$. 
Concretely, $\beta$ and $\alpha$ are proportional to the coefficients in front of the four- and six-derivative effective action, denoted temporarily as $b_4$ and $a_6$ -- And then according to \cite{Cordova:2015vwa} supersymmetry relates them as $b_4\propto \sqrt{a_6}$.
This is therefore also where the analysis in $6d$ and $2d$ works together nicely.

\begin{figure}[t!]
    \begin{subfigure}[t!]{0.5\columnwidth}
        \begin{center}
            \begin{overpic}[ width=0.97\columnwidth]{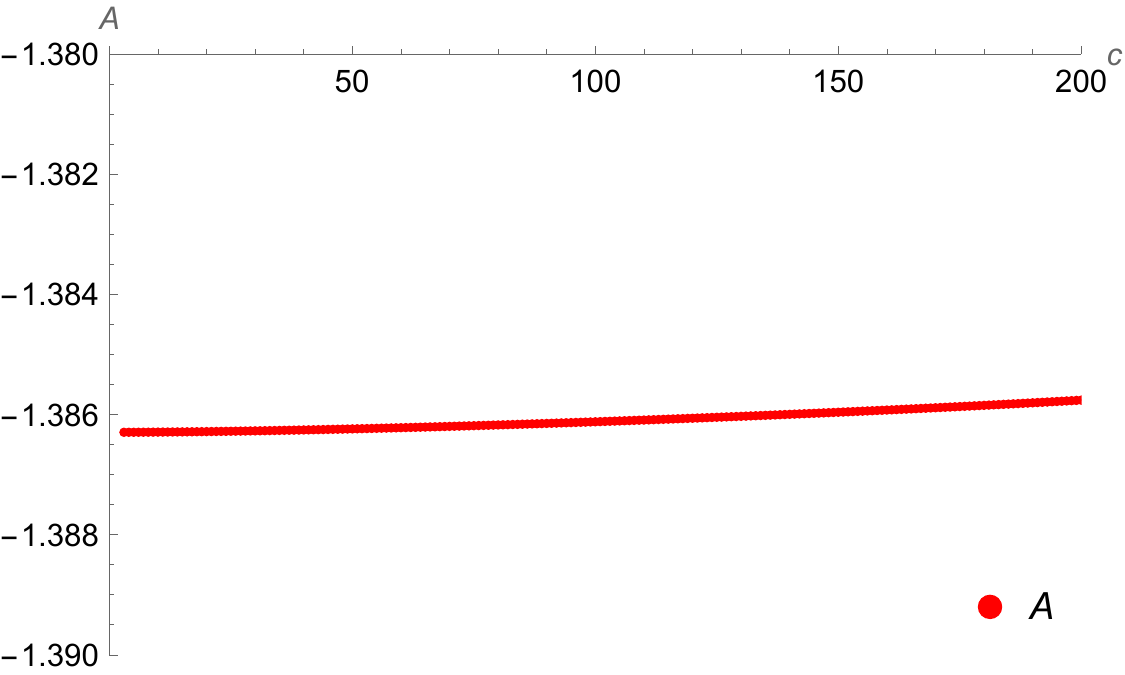}
            \end{overpic}
        \end{center}
    \end{subfigure}
    \begin{subfigure}[t!]{0.5\columnwidth}
        \begin{center}
            \begin{overpic}[width=0.97\columnwidth]{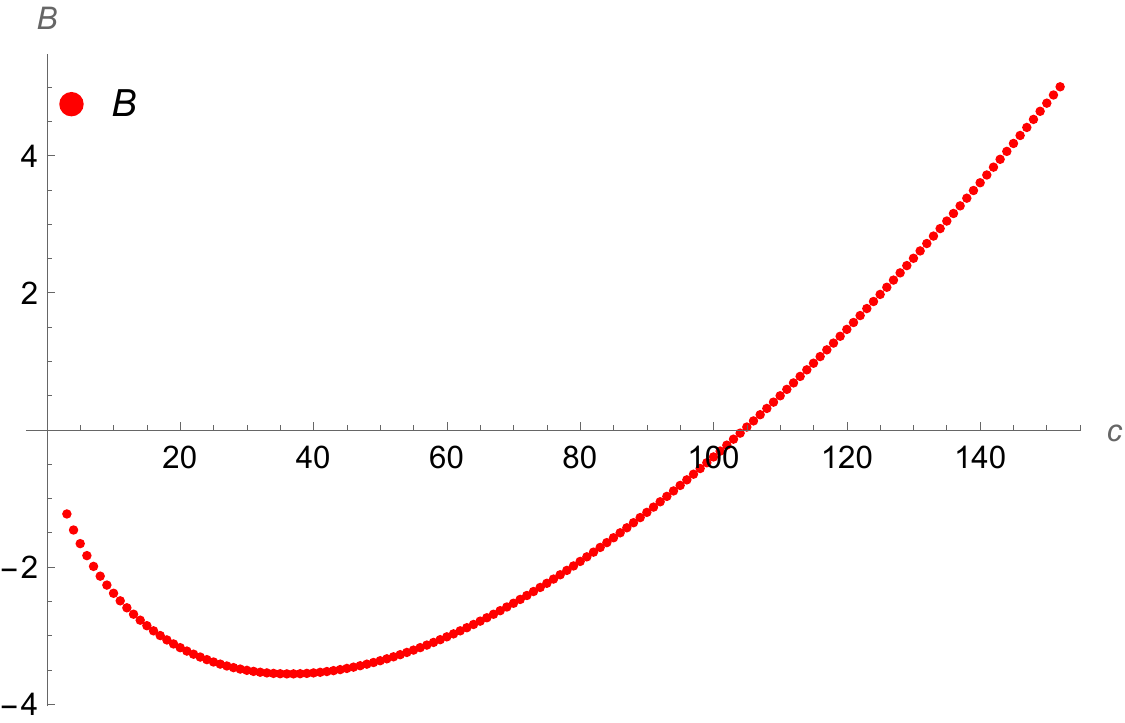}
            \end{overpic}
            \end{center}
    \end{subfigure}
    \caption{(Left) Plot of $A$ for various values of $c$. The data is consistent with $A=-2\log 2$.
    (Right) Plot of $B$ for various values of $c$.}
    \label{fig:AB}
\end{figure}

Assuming all the above we can then determine $A$ and $B$.
Even though they are ambiguous in $6d$, they are still meaningful in the context of $2d$ Virasoro algebra. 
We first subtract the known pieces from $\log \lambda_n$ and define
\begin{align}
    R_n^{[3]}\equiv \log \lambda_n-\left(\log\Gamma(2n+1)+4\sqrt{\frac{c-1}{12}}\sqrt{2n} -\frac{c-1}{12}\log n\right)\;,
\end{align}
and fit it with $\mathtt{A}n+\mathtt{B}+\upgamma n^{-1/2}$, for various values of $n$. 
More precisely, we first fit $\delta R_n^{[3]}$ with $\mathtt{A}-\frac{\upgamma}{2} n^{-3/2}$ to see that
\begin{align}
    A=-1.386\approx -2\log 2,
\end{align}
where the last approximation is only conjectural.
The result of $A$ for various $c$ is shown in Figure \ref{fig:AB}.

Assuming this, we finally subtract the linear in $n$ piece to define
\begin{align}
    R_n^{[4]}\equiv R_n^{[3]}+2n\log 2
\end{align}
and fit it with $\mathtt{B}+\upgamma n^{-1/2}$.
This gives $B$ as a numerical function of $c$, which seems to go as
\begin{align}
    B\sim \frac{c-1}{12}\log\left(\frac{1}{8.0}\frac{c-1}{12}\right)+O(1)
\end{align}
at larger values of $c$, although inconclusive (See Figure \ref{fig:AB}.).
We do not discuss $B$ any further.

\bibliographystyle{JHEP}
\bibliography{refs}

@inproceedings{Witten:1995zh,
	title        = {{Some comments on string dynamics}},
	author       = {Witten, Edward},
	year         = 1995,
	month        = 7,
	booktitle    = {{STRINGS 95: Future Perspectives in String Theory}},
	pages        = {501--523},
	eprint       = {hep-th/9507121},
	archiveprefix = {arXiv},
	reportnumber = {IASSNS-HEP-95-63}
}

@article{Heckman:2020otd,
	title        = {{Qubit Construction in 6D SCFTs}},
	author       = {Heckman, Jonathan J.},
	year         = 2020,
	journal      = {Phys. Lett. B},
	volume       = 811,
	pages        = 135891,
	doi          = {10.1016/j.physletb.2020.135891},
	eprint       = {2007.08545},
	archiveprefix = {arXiv},
	primaryclass = {hep-th}
}

@article{Strominger:1995ac,
	title        = {{Open p-branes}},
	author       = {Strominger, Andrew and Dine, M.},
	year         = 1996,
	journal      = {Phys. Lett. B},
	volume       = 383,
	pages        = {44--47},
	doi          = {10.1201/9781482268737-13},
	eprint       = {hep-th/9512059},
	archiveprefix = {arXiv}
}

@article{Seiberg:1996qx,
	title        = {{Nontrivial fixed points of the renormalization group in six-dimensions}},
	author       = {Seiberg, Nathan},
	year         = 1997,
	journal      = {Phys. Lett. B},
	volume       = 390,
	pages        = {169--171},
	doi          = {10.1016/S0370-2693(96)01424-4},
	eprint       = {hep-th/9609161},
	archiveprefix = {arXiv},
	reportnumber = {RU-96-85}
}

@article{1505.01537,
	title        = {{On the CFT Operator Spectrum at Large Global Charge}},
	author       = {Hellerman, Simeon and Orlando, Domenico and Reffert, Susanne and Watanabe, Masataka},
	year         = 2015,
	journal      = {JHEP},
	volume       = 12,
	pages        = {071},
	doi          = {10.1007/JHEP12(2015)071},
	eprint       = {1505.01537},
	archiveprefix = {arXiv},
	primaryclass = {hep-th}
}

@article{1611.02912,
	title        = {{Semiclassics, Goldstone Bosons and CFT data}},
	author       = {Monin, Alexander and Pirtskhalava, David and Rattazzi, Riccardo and Seibold, Fiona K.},
	year         = 2017,
	journal      = {JHEP},
	volume       = {06},
	pages        = {011},
	doi          = {10.1007/JHEP06(2017)011},
	eprint       = {1611.02912},
	archiveprefix = {arXiv},
	primaryclass = {hep-th}
}

@article{2008.03308,
	title        = {{Selected topics in the large quantum number expansion}},
	author       = {Gaum\'e, Luis \'Alvarez and Orlando, Domenico and Reffert, Susanne},
	year         = 2021,
	journal      = {Phys. Rept.},
	volume       = 933,
	pages        = {1--66},
	doi          = {10.1016/j.physrep.2021.08.001},
	eprint       = {2008.03308},
	archiveprefix = {arXiv},
	primaryclass = {hep-th}
}

@article{1706.05743,
	title        = {{Operator Dimensions from Moduli}},
	author       = {Hellerman, Simeon and Maeda, Shunsuke and Watanabe, Masataka},
	year         = 2017,
	journal      = {JHEP},
	volume       = 10,
	pages        = {089},
	doi          = {10.1007/JHEP10(2017)089},
	eprint       = {1706.05743},
	archiveprefix = {arXiv},
	primaryclass = {hep-th},
	reportnumber = {IPMU-17-0015, CALT-TH-2017-032}
}

@article{1710.07336,
	title        = {{On the Large $R$-charge Expansion in ${\mathcal N} = 2$ Superconformal Field Theories}},
	author       = {Hellerman, Simeon and Maeda, Shunsuke},
	year         = 2017,
	journal      = {JHEP},
	volume       = 12,
	pages        = 135,
	doi          = {10.1007/JHEP12(2017)135},
	eprint       = {1710.07336},
	archiveprefix = {arXiv},
	primaryclass = {hep-th},
	reportnumber = {IPMU17-0143, CALT-TH-2017-059}
}

@article{1804.01535,
	title        = {{Universal correlation functions in rank 1 SCFTs}},
	author       = {Hellerman, Simeon and Maeda, Shunsuke and Orlando, Domenico and Reffert, Susanne and Watanabe, Masataka},
	year         = 2019,
	journal      = {JHEP},
	volume       = 12,
	pages        = {047},
	doi          = {10.1007/JHEP12(2019)047},
	eprint       = {1804.01535},
	archiveprefix = {arXiv},
	primaryclass = {hep-th},
	reportnumber = {CALT-TH-2018-014, IPMU18-0059}
}

@article{1803.00580,
	title        = {{A limit for large $R$-charge correlators in $\mathcal{N}=2$ theories}},
	author       = {Bourget, Antoine and Rodriguez-Gomez, Diego and Russo, Jorge G.},
	year         = 2018,
	journal      = {JHEP},
	volume       = {05},
	pages        = {074},
	doi          = {10.1007/JHEP05(2018)074},
	eprint       = {1803.00580},
	archiveprefix = {arXiv},
	primaryclass = {hep-th}
}

@article{1809.06280,
	title        = {{On the large R-charge $ \mathcal{N} $ = 2 chiral correlators and the Toda equation}},
	author       = {Beccaria, Matteo},
	year         = 2019,
	journal      = {JHEP},
	volume       = {02},
	pages        = {009},
	doi          = {10.1007/JHEP02(2019)009},
	eprint       = {1809.06280},
	archiveprefix = {arXiv},
	primaryclass = {hep-th}
}

@article{1908.10306,
	title        = {{Extremal correlators and random matrix theory}},
	author       = {Grassi, Alba and Komargodski, Zohar and Tizzano, Luigi},
	year         = 2021,
	journal      = {JHEP},
	volume       = {04},
	pages        = 214,
	doi          = {10.1007/JHEP04(2021)214},
	eprint       = {1908.10306},
	archiveprefix = {arXiv},
	primaryclass = {hep-th}
}

@article{2001.06645,
	title        = {{$\mathcal N=2$ conformal gauge theories at large R-charge: the $SU(N)$ case}},
	author       = {Beccaria, Matteo and Galvagno, Francesco and Hasan, Azeem},
	year         = 2020,
	journal      = {JHEP},
	volume       = {03},
	pages        = 160,
	doi          = {10.1007/JHEP03(2020)160},
	eprint       = {2001.06645},
	archiveprefix = {arXiv},
	primaryclass = {hep-th}
}

@article{2005.03021,
	title        = {{S-duality and correlation functions at large R-charge}},
	author       = {Hellerman, Simeon and Maeda, Shunsuke and Orlando, Domenico and Reffert, Susanne and Watanabe, Masataka},
	year         = 2021,
	journal      = {JHEP},
	volume       = {04},
	pages        = 287,
	doi          = {10.1007/JHEP04(2021)287},
	eprint       = {2005.03021},
	archiveprefix = {arXiv},
	primaryclass = {hep-th}
}

@article{2008.01106,
	title        = {{Transition of Large $R$-Charge Operators on a Conformal Manifold}},
	author       = {Sharon, Adar and Watanabe, Masataka},
	year         = 2021,
	journal      = {JHEP},
	volume       = {01},
	pages        = {068},
	doi          = {10.1007/JHEP01(2021)068},
	eprint       = {2008.01106},
	archiveprefix = {arXiv},
	primaryclass = {hep-th}
}

@article{2103.05642,
	title        = {{Large R-charge EFT correlators in N=2 SQCD}},
	author       = {Hellerman, Simeon and Orlando, Domenico},
	year         = 2021,
	month        = 3,
	eprint       = {2103.05642},
	archiveprefix = {arXiv},
	primaryclass = {hep-th}
}

@article{2103.09312,
	title        = {{On the exponentially small corrections to ${\cal N} = 2$ superconformal correlators at large R-charge}},
	author       = {Hellerman, Simeon},
	year         = 2021,
	month        = 3,
	eprint       = {2103.09312},
	archiveprefix = {arXiv},
	primaryclass = {hep-th}
}

@article{2405.19043,
	title        = {{Vacuum Condensates on the Coulomb Branch}},
	author       = {Ivanovskiy, Vyacheslav and Komatsu, Shota and Mishnyakov, Victor and Terziev, Nikolay and Zaigraev, Nikita and Zarembo, Konstantin},
	year         = 2024,
	month        = 5,
	eprint       = {2405.19043},
	archiveprefix = {arXiv},
	primaryclass = {hep-th}
}

@article{2406.19441,
	title        = {{Moduli Spaces in CFT: Large Charge Operators}},
	author       = {Cuomo, Gabriel and Rastelli, Leonardo and Sharon, Adar},
	year         = 2024,
	month        = 6,
	eprint       = {2406.19441},
	archiveprefix = {arXiv},
	primaryclass = {hep-th}
}

@article{2408.07391,
	title        = {{Matrix models for extremal and integrated correlators of higher rank}},
	author       = {Grassi, Alba and Iossa, Cristoforo},
	year         = 2024,
	month        = 8,
	eprint       = {2408.07391},
	archiveprefix = {arXiv},
	primaryclass = {hep-th}
}

@article{Maxfield:2012aw,
	title        = {{The Conformal Anomaly of M5-Branes}},
	author       = {Maxfield, Travis and Sethi, Savdeep},
	year         = 2012,
	journal      = {JHEP},
	volume       = {06},
	pages        = {075},
	doi          = {10.1007/JHEP06(2012)075},
	eprint       = {1204.2002},
	archiveprefix = {arXiv},
	primaryclass = {hep-th}
}

@article{Cordova:2015vwa,
	title        = {{Higher derivative terms, toroidal compactification, and Weyl anomalies in six-dimensional (2, 0) theories}},
	author       = {Cordova, Clay and Dumitrescu, Thomas T. and Yin, Xi},
	year         = 2019,
	journal      = {JHEP},
	volume       = 10,
	pages        = 128,
	doi          = {10.1007/JHEP10(2019)128},
	eprint       = {1505.03850},
	archiveprefix = {arXiv},
	primaryclass = {hep-th}
}

@article{Beem:2014kka,
	title        = {{$ \mathcal{W} $ symmetry in six dimensions}},
	author       = {Beem, Christopher and Rastelli, Leonardo and van Rees, Balt C.},
	year         = 2015,
	journal      = {JHEP},
	volume       = {05},
	pages        = {017},
	doi          = {10.1007/JHEP05(2015)017},
	eprint       = {1404.1079},
	archiveprefix = {arXiv},
	primaryclass = {hep-th},
	reportnumber = {CERN-PH-TH-2014-056, CERN-PH-TH-2014-56}
}

@article{Chen:2017yze,
	title        = {{A Numerical Approach to Virasoro Blocks and the Information Paradox}},
	author       = {Chen, Hongbin and Hussong, Charles and Kaplan, Jared and Li, Daliang},
	year         = 2017,
	journal      = {JHEP},
	volume       = {09},
	pages        = 102,
	doi          = {10.1007/JHEP09(2017)102},
	eprint       = {1703.09727},
	archiveprefix = {arXiv},
	primaryclass = {hep-th}
}

@article{Nahm:1977tg,
	title        = {{Supersymmetries and Their Representations}},
	author       = {Nahm, W.},
	year         = 1978,
	journal      = {Nucl. Phys. B},
	volume       = 135,
	pages        = 149,
	doi          = {10.1201/9781482268737-2},
	reportnumber = {CERN-TH-2341}
}

@article{hep-th/9707079,
	title        = {{Matrix description of interacting theories in six-dimensions}},
	author       = {Aharony, Ofer and Berkooz, M. and Kachru, S. and Seiberg, N. and Silverstein, Eva},
	year         = 1998,
	journal      = {Adv. Theor. Math. Phys.},
	volume       = 1,
	pages        = {148--157},
	doi          = {10.4310/ATMP.1997.v1.n1.a5},
	eprint       = {hep-th/9707079},
	archiveprefix = {arXiv},
	reportnumber = {IASSNS-HEP-97-74, RU-97-74}
}

@article{hep-th/0702069,
	title        = {{Supersymmetric states in M5/M2 CFTs}},
	author       = {Bhattacharyya, Sayantani and Minwalla, Shiraz},
	year         = 2007,
	journal      = {JHEP},
	volume       = 12,
	pages        = {004},
	doi          = {10.1088/1126-6708/2007/12/004},
	eprint       = {hep-th/0702069},
	archiveprefix = {arXiv}
}

@article{1205.3994,
	title        = {{On renormalization group flows and the a-theorem in 6d}},
	author       = {Elvang, Henriette and Freedman, Daniel Z. and Hung, Ling-Yan and Kiermaier, Michael and Myers, Robert C. and Theisen, Stefan},
	year         = 2012,
	journal      = {JHEP},
	volume       = 10,
	pages        = {011},
	doi          = {10.1007/JHEP10(2012)011},
	eprint       = {1205.3994},
	archiveprefix = {arXiv},
	primaryclass = {hep-th},
	reportnumber = {MCTP-12-09, MIT-CTP-4362, PUPT-2413, SU-ITP-12-14}
}

@article{hep-th/0001041,
	title        = {{Conformal anomaly of (2,0) tensor multiplet in six-dimensions and AdS / CFT correspondence}},
	author       = {Bastianelli, F. and Frolov, S. and Tseytlin, Arkady A.},
	year         = 2000,
	journal      = {JHEP},
	volume       = {02},
	pages        = {013},
	doi          = {10.1088/1126-6708/2000/02/013},
	eprint       = {hep-th/0001041},
	archiveprefix = {arXiv},
	reportnumber = {UAHEP-00-1, OHSTPY-HEP-T-00-001}
}

@article{1912.09479,
	title        = {{Renormalization Group Flows, the $a$-Theorem and Conformal Bootstrap}},
	author       = {Kundu, Sandipan},
	year         = 2020,
	journal      = {JHEP},
	volume       = {05},
	pages        = {014},
	doi          = {10.1007/JHEP05(2020)014},
	eprint       = {1912.09479},
	archiveprefix = {arXiv},
	primaryclass = {hep-th}
}

@article{2012.10450,
	title        = {{RG flows with global symmetry breaking and bounds from chaos}},
	author       = {Kundu, Sandipan},
	year         = 2022,
	journal      = {Phys. Rev. D},
	volume       = 105,
	number       = 2,
	pages        = {025016},
	doi          = {10.1103/PhysRevD.105.025016},
	eprint       = {2012.10450},
	archiveprefix = {arXiv},
	primaryclass = {hep-th}
}

@article{2103.13395,
	title        = {{Effective field theory of 6D SUSY RG Flows}},
	author       = {Heckman, Jonathan J. and Kundu, Sandipan and Zhang, Hao Y.},
	year         = 2021,
	journal      = {Phys. Rev. D},
	volume       = 104,
	number       = 8,
	pages        = {085017},
	doi          = {10.1103/PhysRevD.104.085017},
	eprint       = {2103.13395},
	archiveprefix = {arXiv},
	primaryclass = {hep-th}
}

@article{1610.04495,
	title        = {{Compensating strong coupling with large charge}},
	author       = {Alvarez-Gaume, Luis and Loukas, Orestis and Orlando, Domenico and Reffert, Susanne},
	year         = 2017,
	journal      = {JHEP},
	volume       = {04},
	pages        = {059},
	doi          = {10.1007/JHEP04(2017)059},
	eprint       = {1610.04495},
	archiveprefix = {arXiv},
	primaryclass = {hep-th},
	reportnumber = {CERN-TH-2016-221}
}

@article{1705.05825,
	title        = {{A Note on Inhomogeneous Ground States at Large Global Charge}},
	author       = {Hellerman, Simeon and Kobayashi, Nozomu and Maeda, Shunsuke and Watanabe, Masataka},
	year         = 2019,
	journal      = {JHEP},
	volume       = 10,
	pages        = {038},
	doi          = {10.1007/JHEP10(2019)038},
	eprint       = {1705.05825},
	archiveprefix = {arXiv},
	primaryclass = {hep-th},
	reportnumber = {IPMU17-0081, CALT-TH-2017-24}
}

@article{1804.06495,
	title        = {{Observables in inhomogeneous ground states at large global charge}},
	author       = {Hellerman, Simeon and Kobayashi, Nozomu and Maeda, Shunsuke and Watanabe, Masataka},
	year         = 2021,
	journal      = {JHEP},
	volume       = {08},
	pages        = {079},
	doi          = {10.1007/JHEP08(2021)079},
	eprint       = {1804.06495},
	archiveprefix = {arXiv},
	primaryclass = {hep-th},
	reportnumber = {IPMU18-0069}
}

@article{1904.09815,
	title        = {{Chern-Simons-matter theories at large baryon number}},
	author       = {Watanabe, Masataka},
	year         = 2021,
	journal      = {JHEP},
	volume       = 10,
	pages        = 245,
	doi          = {10.1007/JHEP10(2021)245},
	eprint       = {1904.09815},
	archiveprefix = {arXiv},
	primaryclass = {hep-th}
}

@article{1909.02571,
	title        = {{Large charge at large N}},
	author       = {Alvarez-Gaume, Luis and Orlando, Domenico and Reffert, Susanne},
	year         = 2019,
	journal      = {JHEP},
	volume       = 12,
	pages        = 142,
	doi          = {10.1007/JHEP12(2019)142},
	eprint       = {1909.02571},
	archiveprefix = {arXiv},
	primaryclass = {hep-th}
}

@article{2203.08843,
	title        = {{Stability analysis of a non-unitary CFT}},
	author       = {Watanabe, Masataka},
	year         = 2023,
	journal      = {JHEP},
	volume       = 11,
	pages        = {042},
	doi          = {10.1007/JHEP11(2023)042},
	eprint       = {2203.08843},
	archiveprefix = {arXiv},
	primaryclass = {hep-th}
}

@article{2003.13121,
	title        = {{Charging the $O(N)$ model}},
	author       = {Antipin, Oleg and Bersini, Jahmall and Sannino, Francesco and Wang, Zhi-Wei and Zhang, Chen},
	year         = 2020,
	journal      = {Phys. Rev. D},
	volume       = 102,
	number       = 4,
	pages        = {045011},
	doi          = {10.1103/PhysRevD.102.045011},
	eprint       = {2003.13121},
	archiveprefix = {arXiv},
	primaryclass = {hep-th},
	reportnumber = {RBI-ThPhys-2020-09, CP3-Origins-2020-04 DNRF90, NCTS-TH/2005}
}

@article{Cordova:2016emh,
	title        = {{Multiplets of Superconformal Symmetry in Diverse Dimensions}},
	author       = {Cordova, Clay and Dumitrescu, Thomas T. and Intriligator, Kenneth},
	year         = 2019,
	journal      = {JHEP},
	volume       = {03},
	pages        = 163,
	doi          = {10.1007/JHEP03(2019)163},
	eprint       = {1612.00809},
	archiveprefix = {arXiv},
	primaryclass = {hep-th}
}

@article{1612.08985,
	title        = {{Abelian scalar theory at large global charge}},
	author       = {Loukas, Orestis},
	year         = 2017,
	journal      = {Fortsch. Phys.},
	volume       = 65,
	number       = 9,
	pages        = 1700028,
	doi          = {10.1002/prop.201700028},
	eprint       = {1612.08985},
	archiveprefix = {arXiv},
	primaryclass = {hep-th}
}

@article{2103.16580,
	title        = {{D-branes and orbit average}},
	author       = {Yang, Peihe and Jiang, Yunfeng and Komatsu, Shota and Wu, Jun-Bao},
	year         = 2022,
	journal      = {SciPost Phys.},
	volume       = 12,
	number       = 2,
	pages        = {055},
	doi          = {10.21468/SciPostPhys.12.2.055},
	eprint       = {2103.16580},
	archiveprefix = {arXiv},
	primaryclass = {hep-th},
	reportnumber = {CERN-TH-2021-043, USTC-ICTS/PCFT-21-15}
}

@article{Cordova:2016xhm,
	title        = {{Deformations of Superconformal Theories}},
	author       = {Cordova, Clay and Dumitrescu, Thomas T. and Intriligator, Kenneth},
	year         = 2016,
	journal      = {JHEP},
	volume       = 11,
	pages        = 135,
	doi          = {10.1007/JHEP11(2016)135},
	eprint       = {1602.01217},
	archiveprefix = {arXiv},
	primaryclass = {hep-th}
}

@article{Beem:2013sza,
	title        = {{Infinite Chiral Symmetry in Four Dimensions}},
	author       = {Beem, Christopher and Lemos, Madalena and Liendo, Pedro and Peelaers, Wolfger and Rastelli, Leonardo and van Rees, Balt C.},
	year         = 2015,
	journal      = {Commun. Math. Phys.},
	volume       = 336,
	number       = 3,
	pages        = {1359--1433},
	doi          = {10.1007/s00220-014-2272-x},
	eprint       = {1312.5344},
	archiveprefix = {arXiv},
	primaryclass = {hep-th},
	reportnumber = {YITP-SB-13-45, CERN-PH-TH-2013-311, HU-EP-13-78}
}

@article{1502.07742,
	title        = {{Virasoro conformal blocks in closed form}},
	author       = {Perlmutter, Eric},
	year         = 2015,
	journal      = {JHEP},
	volume       = {08},
	pages        = {088},
	doi          = {10.1007/JHEP08(2015)088},
	eprint       = {1502.07742},
	archiveprefix = {arXiv},
	primaryclass = {hep-th}
}

@article{1710.10458,
	title        = {{Universality of Quantum Information in Chaotic CFTs}},
	author       = {Lashkari, Nima and Dymarsky, Anatoly and Liu, Hong},
	year         = 2018,
	journal      = {JHEP},
	volume       = {03},
	pages        = {070},
	doi          = {10.1007/JHEP03(2018)070},
	eprint       = {1710.10458},
	archiveprefix = {arXiv},
	primaryclass = {hep-th}
}

@article{1501.05315,
	title        = {{Virasoro Conformal Blocks and Thermality from Classical Background Fields}},
	author       = {Fitzpatrick, A. Liam and Kaplan, Jared and Walters, Matthew T.},
	year         = 2015,
	journal      = {JHEP},
	volume       = 11,
	pages        = 200,
	doi          = {10.1007/JHEP11(2015)200},
	eprint       = {1501.05315},
	archiveprefix = {arXiv},
	primaryclass = {hep-th}
}

@article{1510.00014,
	title        = {{Hawking from Catalan}},
	author       = {Fitzpatrick, A. Liam and Kaplan, Jared and Walters, Matthew T. and Wang, Junpu},
	year         = 2016,
	journal      = {JHEP},
	volume       = {05},
	pages        = {069},
	doi          = {10.1007/JHEP05(2016)069},
	eprint       = {1510.00014},
	archiveprefix = {arXiv},
	primaryclass = {hep-th}
}

@article{1609.07153,
	title        = {{On the Late-Time Behavior of Virasoro Blocks and a Classification of Semiclassical Saddles}},
	author       = {Fitzpatrick, A. Liam and Kaplan, Jared},
	year         = 2017,
	journal      = {JHEP},
	volume       = {04},
	pages        = {072},
	doi          = {10.1007/JHEP04(2017)072},
	eprint       = {1609.07153},
	archiveprefix = {arXiv},
	primaryclass = {hep-th}
}

@article{1804.06171,
	title        = {{New Properties of Large-$c$ Conformal Blocks from Recursion Relation}},
	author       = {Kusuki, Yuya},
	year         = 2018,
	journal      = {JHEP},
	volume       = {07},
	pages        = {010},
	doi          = {10.1007/JHEP07(2018)010},
	eprint       = {1804.06171},
	archiveprefix = {arXiv},
	primaryclass = {hep-th},
	reportnumber = {YITP-18-22}
}

@article{1806.04352,
	title        = {{Large $c$ Virasoro Blocks from Monodromy Method beyond Known Limits}},
	author       = {Kusuki, Yuya},
	year         = 2018,
	journal      = {JHEP},
	volume       = {08},
	pages        = 161,
	doi          = {10.1007/JHEP08(2018)161},
	eprint       = {1806.04352},
	archiveprefix = {arXiv},
	primaryclass = {hep-th},
	reportnumber = {YITP-1863}
}

@article{1910.04169,
	title        = {{Semi-classical Virasoro blocks: proof of exponentiation}},
	author       = {Be\c{s}ken, Mert and Datta, Shouvik and Kraus, Per},
	year         = 2020,
	journal      = {JHEP},
	volume       = {01},
	pages        = 109,
	doi          = {10.1007/JHEP01(2020)109},
	eprint       = {1910.04169},
	archiveprefix = {arXiv},
	primaryclass = {hep-th}
}

@article{2007.10998,
	title        = {{Virasoro blocks and quasimodular forms}},
	author       = {Das, Diptarka and Datta, Shouvik and Raman, Madhusudhan},
	year         = 2020,
	journal      = {JHEP},
	volume       = 11,
	pages        = {010},
	doi          = {10.1007/JHEP11(2020)010},
	eprint       = {2007.10998},
	archiveprefix = {arXiv},
	primaryclass = {hep-th},
	reportnumber = {CERN-TH-2020-119}
}

@article{Cardona:2020cfy,
	title        = {{Virasoro blocks at large exchange dimension}},
	author       = {Cardona, Carlos},
	year         = 2021,
	journal      = {Nucl. Phys. B},
	volume       = 963,
	pages        = 115284,
	doi          = {10.1016/j.nuclphysb.2020.115284},
	eprint       = {2006.01237},
	archiveprefix = {arXiv},
	primaryclass = {hep-th}
}

@article{1711.09913,
	title        = {{Renyi entropy for local quenches in 2D CFT from numerical conformal blocks}},
	author       = {Kusuki, Yuya and Takayanagi, Tadashi},
	year         = 2018,
	journal      = {JHEP},
	volume       = {01},
	pages        = 115,
	doi          = {10.1007/JHEP01(2018)115},
	eprint       = {1711.09913},
	archiveprefix = {arXiv},
	primaryclass = {hep-th},
	reportnumber = {YITP-17-126, IPMU17-0163}
}

@article{1810.01335,
	title        = {{Light Cone Bootstrap in General 2D CFTs and Entanglement from Light Cone Singularity}},
	author       = {Kusuki, Yuya},
	year         = 2019,
	journal      = {JHEP},
	volume       = {01},
	pages        = {025},
	doi          = {10.1007/JHEP01(2019)025},
	eprint       = {1810.01335},
	archiveprefix = {arXiv},
	primaryclass = {hep-th},
	reportnumber = {YITP-18-106}
}

@article{1905.02191,
	title        = {{Entanglement Entropy, OTOC and Bootstrap in 2D CFTs from Regge and Light Cone Limits of Multi-point Conformal Block}},
	author       = {Kusuki, Yuya and Miyaji, Masamichi},
	year         = 2019,
	journal      = {JHEP},
	volume       = {08},
	pages        = {063},
	doi          = {10.1007/JHEP08(2019)063},
	eprint       = {1905.02191},
	archiveprefix = {arXiv},
	primaryclass = {hep-th},
	reportnumber = {YITP-19-30 UT-19-08}
}

@inproceedings{Beemtalk,
	title        = {{VOAs and 4d SCFTs (part II)}},
	author       = {Beem, Chris},
	year         = 2018,
	booktitle    = {Talk at SCGP workshop on Vertex Algebras and Gauge Theory}
}

@article{Beem:2017ooy,
	title        = {{Vertex operator algebras, Higgs branches, and modular differential equations}},
	author       = {Beem, Christopher and Rastelli, Leonardo},
	year         = 2018,
	journal      = {JHEP},
	volume       = {08},
	pages        = 114,
	doi          = {10.1007/JHEP08(2018)114},
	eprint       = {1707.07679},
	archiveprefix = {arXiv},
	primaryclass = {hep-th},
	reportnumber = {YITP-SB-17-27}
}

@article{Gunaydin:1984fk,
	title        = {{The Spectrum of the $S^5$ Compactification of the Chiral $\mathcal{N}=2$, $D=10$ Supergravity and the Unitary Supermultiplets of $U(2, 2|4)$}},
	author       = {Gunaydin, M. and Marcus, N.},
	year         = 1985,
	journal      = {Class. Quant. Grav.},
	volume       = 2,
	pages        = {L11},
	doi          = {10.1088/0264-9381/2/2/001},
	reportnumber = {CALT-68-1185}
}

@article{Graham:1999jg,
	title        = {{Volume and area renormalizations for conformally compact Einstein metrics}},
	author       = {Graham, C. Robin},
	year         = 2000,
	journal      = {Rend. Circ. Mat. Palermo S},
	volume       = 63,
	pages        = {31--42},
	editor       = {Slov\'ak, Jan and Cadek, Martin},
	eprint       = {math/9909042},
	archiveprefix = {arXiv}
}

@article{Intriligator:2000eq,
	title        = {{Anomaly matching and a Hopf-Wess-Zumino term in $6d$, $\mathcal{N}=(2,0)$ field theories}},
	author       = {Intriligator, Kenneth A.},
	year         = 2000,
	journal      = {Nucl. Phys. B},
	volume       = 581,
	pages        = {257--273},
	doi          = {10.1016/S0550-3213(00)00148-6},
	eprint       = {hep-th/0001205},
	archiveprefix = {arXiv},
	reportnumber = {UCSD-PTH-00-02, IASSNS-HEP-00-06, RUNHETC-00-03}
}

@article{Henningson:1998gx,
	title        = {{The Holographic Weyl anomaly}},
	author       = {Henningson, M. and Skenderis, K.},
	year         = 1998,
	journal      = {JHEP},
	volume       = {07},
	pages        = {023},
	doi          = {10.1088/1126-6708/1998/07/023},
	eprint       = {hep-th/9806087},
	archiveprefix = {arXiv},
	reportnumber = {CERN-TH-98-188, KUL-TF-98-21}
}

@article{Capper:1974ic,
	title        = {{Trace anomalies in dimensional regularization}},
	author       = {Capper, D. M. and Duff, M. J.},
	year         = 1974,
	journal      = {Nuovo Cim. A},
	volume       = 23,
	pages        = {173--183},
	doi          = {10.1007/BF02748300}
}

@article{Gunaydin:1985tc,
	title        = {{Unitary Supermultiplets of $Osp(8|4,\mathbb{R})$ and the Spectrum of the $S^7$ Compactification of Eleven-dimensional Supergravity}},
	author       = {Gunaydin, M. and Warner, N. P.},
	year         = 1986,
	journal      = {Nucl. Phys. B},
	volume       = 272,
	pages        = {99--124},
	doi          = {10.1016/0550-3213(86)90342-1},
	reportnumber = {CALT-68-1202}
}

@article{Gunaydin:1984wc,
	title        = {{General Construction of the Unitary Representations of Anti-de Sitter Superalgebras and the Spectrum of the $S^4$ Compactification of Eleven-dimensional Supergravity}},
	author       = {Gunaydin, M. and van Nieuwenhuizen, P. and Warner, N. P.},
	year         = 1985,
	journal      = {Nucl. Phys. B},
	volume       = 255,
	pages        = {63--92},
	doi          = {10.1016/0550-3213(85)90129-4},
	reportnumber = {CALT-68-1184}
}

@article{Hanany:2000fq,
	title        = {{On orientifolds, discrete torsion, branes and M theory}},
	author       = {Hanany, Amihay and Kol, Barak},
	year         = 2000,
	journal      = {JHEP},
	volume       = {06},
	pages        = {013},
	doi          = {10.1088/1126-6708/2000/06/013},
	eprint       = {hep-th/0003025},
	archiveprefix = {arXiv},
	reportnumber = {MIT-CTP-2946, TAUP-2609-2000, NSF-ITP-00-07}
}

@article{Heckman:2018jxk,
	title        = {{Top Down Approach to 6D SCFTs}},
	author       = {Heckman, Jonathan J. and Rudelius, Tom},
	year         = 2019,
	journal      = {J. Phys. A},
	volume       = 52,
	number       = 9,
	pages        = {093001},
	doi          = {10.1088/1751-8121/aafc81},
	eprint       = {1805.06467},
	archiveprefix = {arXiv},
	primaryclass = {hep-th}
}

@article{Argyres:2022mnu,
	title        = {{Snowmass White Paper on SCFTs}},
	author       = {Argyres, Philip C. and Heckman, Jonathan J. and Intriligator, Kenneth and Martone, Mario},
	year         = 2022,
	month        = 2,
	eprint       = {2202.07683},
	archiveprefix = {arXiv},
	primaryclass = {hep-th}
}

@article{1986JETP...63.1061Z,
	title        = {{Two-dimensional conformal symmetry and critical four-spin correlation functions in the Ashkin-Teller model}},
	author       = {{Zamolodchikov}, Al. B.},
	year         = 1986,
	month        = may,
	journal      = {Soviet Journal of Experimental and Theoretical Physics},
	volume       = 63,
	number       = 5,
	pages        = 1061,
	adsurl       = {https://ui.adsabs.harvard.edu/abs/1986JETP...63.1061Z}
}

@article{cmp/1103941860,
	title        = {{Conformal symmetry in two dimensions: an explicit recurrence formula for the conformal partial wave amplitude}},
	author       = {Al. B. Zamolodchikov},
	year         = 1984,
	journal      = {Communications in Mathematical Physics},
	publisher    = {Springer},
	volume       = 96,
	number       = 3,
	pages        = {419 -- 422}
}

@article{2007.07262,
	title        = {{6D SCFTs, 4D SCFTs, Conformal Matter, and Spin Chains}},
	author       = {Baume, Florent and Heckman, Jonathan J. and Lawrie, Craig},
	year         = 2021,
	journal      = {Nucl. Phys. B},
	volume       = 967,
	pages        = 115401,
	doi          = {10.1016/j.nuclphysb.2021.115401},
	eprint       = {2007.07262},
	archiveprefix = {arXiv},
	primaryclass = {hep-th}
}

@article{2208.02272,
	title        = {{Super-spin chains for 6D SCFTs}},
	author       = {Baume, Florent and Heckman, Jonathan J. and Lawrie, Craig},
	year         = 2023,
	journal      = {Nucl. Phys. B},
	volume       = 992,
	pages        = 116250,
	doi          = {10.1016/j.nuclphysb.2023.116250},
	eprint       = {2208.02272},
	archiveprefix = {arXiv},
	primaryclass = {hep-th},
	reportnumber = {DESY-22-132}
}

@article{hep-th/0202021,
	title        = {{Strings in flat space and pp waves from N=4 superYang-Mills}},
	author       = {Berenstein, David Eliecer and Maldacena, Juan Martin and Nastase, Horatiu Stefan},
	year         = 2002,
	journal      = {JHEP},
	volume       = {04},
	pages        = {013},
	doi          = {10.1088/1126-6708/2002/04/013},
	eprint       = {hep-th/0202021},
	archiveprefix = {arXiv}
}

@article{Zamolodchikov:1987avt,
	title        = {{Conformal symmetry in two-dimensional space: Recursion representation of conformal block}},
	author       = {Zamolodchikov, Al. B.},
	year         = 1987,
	journal      = {Theor. Math. Phys.},
	volume       = 73,
	number       = 1,
	pages        = {1088--1093},
	doi          = {10.1007/BF01022967}
}

@article{Bergman:2020bvi,
	title        = {{Charges and holography in 6d (1,0) theories}},
	author       = {Bergman, Oren and Fazzi, Marco and Rodr\'\i{}guez-G\'omez, Diego and Tomasiello, Alessandro},
	year         = 2020,
	journal      = {JHEP},
	volume       = {05},
	pages        = 138,
	doi          = {10.1007/JHEP05(2020)138},
	eprint       = {2002.04036},
	archiveprefix = {arXiv},
	primaryclass = {hep-th}
}

@phdthesis{Thielemans:1994er,
	title        = {{An Algorithmic approach to operator product expansions, W algebras and W strings}},
	author       = {Thielemans, Kris},
	year         = 1994,
	eprint       = {hep-th/9506159},
	archiveprefix = {arXiv},
	school       = {Leuven U.}
}

@article{Watanabe:2019pdh,
	title        = {{Accessing large global charge via the $\epsilon$-expansion}},
	author       = {Watanabe, Masataka},
	year         = 2021,
	journal      = {JHEP},
	volume       = {04},
	pages        = 264,
	doi          = {10.1007/JHEP04(2021)264},
	eprint       = {1909.01337},
	archiveprefix = {arXiv},
	primaryclass = {hep-th}
}

@article{Arias-Tamargo:2019xld,
	title        = {{The large charge limit of scalar field theories and the Wilson-Fisher fixed point at $\epsilon=0$}},
	author       = {Arias-Tamargo, G. and Rodriguez-Gomez, D. and Russo, J. G.},
	year         = 2019,
	journal      = {JHEP},
	volume       = 10,
	pages        = 201,
	doi          = {10.1007/JHEP10(2019)201},
	eprint       = {1908.11347},
	archiveprefix = {arXiv},
	primaryclass = {hep-th}
}

@article{Badel:2019oxl,
	title        = {{The Epsilon Expansion Meets Semiclassics}},
	author       = {Badel, Gil and Cuomo, Gabriel and Monin, Alexander and Rattazzi, Riccardo},
	year         = 2019,
	journal      = {JHEP},
	volume       = 11,
	pages        = 110,
	doi          = {10.1007/JHEP11(2019)110},
	eprint       = {1909.01269},
	archiveprefix = {arXiv},
	primaryclass = {hep-th}
}

\end{document}